\documentclass[ejs]{imsart}

\usepackage{graphicx}
\usepackage{latexsym,amsmath,amssymb,amsthm,epic,eepic,multirow}
\usepackage{natbib}

\usepackage{graphicx}
\usepackage{multirow}
\usepackage{subfigure}
\usepackage{natbib}
\usepackage{algorithm}
\usepackage{algorithmic}

\usepackage{tikz}
\usetikzlibrary{decorations.markings}
\tikzstyle{vertex}=[circle, draw, fill, inner sep=0pt, minimum size=0.15cm]
\usetikzlibrary{arrows,automata,decorations.markings}
\usetikzlibrary{patterns,snakes}
\usetikzlibrary{positioning}
\newcommand{\vertex}{\node[vertex]}

\startlocaldefs

\newcommand{\PP}{\mathbb{P}}
\newcommand{\EE}{\mathbb{E}}
\newcommand{\R}{\mathbb{R}}

\newcommand{\eps}{\varepsilon}

\newcommand{\pa}{\mathrm{pa}}
\newcommand{\Do}{\mathrm{do}}

\newcommand{\argminn}{\operatornamewithlimits{argmin}}

\newtheorem{theo}{Theorem}
\newtheorem{prop}{Proposition}

\newtheorem{rem}{Remark}
\newtheorem{assu}{Assumption}

\newtheorem{corr}{Corollary}
\newenvironment{prf}{{Proof. }}{\qed}

\usepackage{color}

\endlocaldefs

\begin{document}

\begin{frontmatter}


\title{Marginal integration for\\
nonparametric causal inference}

\runtitle{Marginal integration for nonparametric causal inference}

\author{\fnms{Jan}
  \snm{Ernest}\thanksref{t1}\corref{}\ead[label=e1]{ernest@stat.math.ethz.ch}
  \and \fnms{Peter}
  \snm{B\"uhlmann}\ead[label=e3]{buhlmann@stat.math.ethz.ch}} 
\address{Seminar f\"ur Statistik \\ ETH Z\"urich}
\address{\printead{e1,e3}}
\thankstext{t1}{Partially supported by the Swiss National Science
  Foundation grant no. 20PA20E-134493.}

\runauthor{Ernest and B\"uhlmann}

\begin{abstract}
We consider the problem of inferring the total causal effect of a single
continuous variable intervention on a (response) variable of interest. We propose a
certain marginal integration regression technique for a very general class
of potentially nonlinear structural equation models (SEMs) with known
structure, or at least known superset of adjustment 
variables: we call the procedure \emph{S-mint} regression. We easily
  derive that it achieves the convergence rate as for nonparametric regression: for example, single variable intervention effects can be estimated with convergence rate $n^{-2/5}$ assuming smoothness with twice differentiable functions.  
Our result can also be seen as a major robustness property with respect 
to model misspecification which goes much beyond the notion of double
robustness. Furthermore, when the structure of the SEM is 
not known, we can estimate (the equivalence class of) the directed acyclic graph
corresponding to the SEM, and then proceed by using \emph{S-mint} based on
these estimates. 
We empirically compare the \emph{S-mint} regression method 
with more classical approaches and argue that the former is indeed more
robust, more reliable and substantially simpler. 
\end{abstract}





\end{frontmatter}

\section{Introduction}
Understanding cause-effect relationships between variables is of
great interest in many fields of science. An ambitious but highly desirable
goal is to infer causal effects from observational data obtained by
observing a system of interest without subjecting it to
interventions.\footnote{More generally, in the presence of both, interventional
  and observational data, the goal is to infer intervention or causal
  effects among variables which are not directly targeted by the
  interventions from interventional data.} This would allow to circumvent
potentially severe experimental constraints or to substantially lower
experimental costs. The words ``causal inference'' (usually) refer to the
problem of inferring effects which are due to (or caused by) interventions: if
we make an outside intervention at a variable~$X$, say, what is its effect
on another response variable of interest~$Y$. We describe 
examples in
Section~\ref{subsec.applic}. 
Various fields and concepts have contributed to the
understanding and quantification of causal inference: the framework of
potential outcomes and counterfactuals \citep[cf.][]{rubin2005causal},
see also \citet{dawid2000causal}, structural equation
modeling \citep[cf.][]{bollen1998structural}, and graphical modeling
\citep[cf.][]{lauritzen1988local,greenland1999causal}; the book by \citet{pearl00} provides a nice overview.  

We consider aspects of the problem indicated above, namely inferring
intervention or causal 
effects from observational data without external interventions. Thus, we 
deal (in part) with the question of how to infer causal effects
\emph{without} relying  
on randomized experiments or randomized studies. Besides fundamental
conceptual aspects, as 
treated for example in the books by \citet{pearl00}, \citet{sgs00} and
\citet{koller2009probabilistic}, important issues include statistical tasks
such as estimation accuracy and 
robustness with respect to model misspecification. This paper focuses on
the two latter topics, covering also high-dimensional sparse settings with many
variables (parameters) but relatively few observational data points. 

In general, the tools for inferring causal effects are
different from regression methods, but as we will argue, the regression methods, when properly applied, remain a useful tool for causal inference. In fact, for the estimation of total causal effects, we make use of a marginal integration regression method which has originally been
proposed for additive regression modeling \citep{linton1995kernel}. Its use
in causal inference is novel. Relying on known theory for marginal
integration in regression \citep{fan1998direct}, our main result
(Theorem~\ref{th1}) establishes optimal convergence properties and
justifies the method as a fully 
robust procedure against model misspecification, as explained further in
Section~\ref{subsec.ourcontr}.  

\subsection{Basic concepts and definitions for causal
  inference}\label{subsec.basic}

We very briefly introduce some of the basic concepts for causal
inference (and the reader who is familiar with them can skip this
subsection). We consider $p$ random variables $X_1,\ldots ,X_p$, where one of
them is a response variable $Y$ of interest and one of them an intervention
variable $X$, that is, the variable where we make an external intervention
by setting $X$ to a certain value $x$. Such an intervention is denoted by
Pearl's do-operator $\Do(X=x)$ \citep[cf.][]{pearl00}. We denote
the indices corresponding to $Y$ and $X$ by $j_Y$ and $j_X$, respectively:
thus, $Y = X_{j_Y}$ and $X = X_{j_X}$. We assume a setting where all
relevant variables are observed, i.e., there are no relevant hidden
variables.\footnote{It suffices to assume that $Y$, $X$ and $X_{\pa(j_X)}$
  (the parents of $X$) are observed, see (\ref{backdoor0}).}

The system of variables is assumed to be generated from a structural equation model (SEM): 
\begin{align}\label{SEM0}
X_j \leftarrow f_{j}(X_{\pa(j)},\eps_j),\ j=1,\ldots ,p.
\end{align}
Thereby, $\eps_1,\ldots ,\eps_p$ are independent noise (or innovation)
variables, and there is an underlying structure in terms of a directed
acyclic graph (DAG) $D$, where each node $j$ corresponds to the random
variable $X_j$: We denote by $\pa(j) = \pa_D(j)$ the set of parents of node
$j$ in the underlying DAG $D$,\footnote{The set of 
  parents is $\pa_D(j) =  \{k;\ \mbox{there exists a directed edge $k \to
    j$ in DAG $D$}\}$.} and $f_j(\cdot)$ are assumed to be real-valued
(measurable) functions. For any index set $U \subseteq
\{1,\ldots ,p\}$ we write $X_U := (X_v)_{v \in U}$, for example, $X_{\pa(j)} = (X_v)_{v \in \pa(j)}$.

The causal mechanism we are interested in is the
total effect of an intervention at a single variable $X$ on a response
variable $Y$ of interest.\footnote{A total effect is the effect of an
  intervention at a variable $X$ to another variable $Y$, taking into account the total of all (directed) paths from $X$ to $Y$.} The distribution
of $Y$ when doing an external 
intervention $\Do(X=x)$ by setting variable $X$ to $x$ is 
identified with its
density (assumed to exist) or discrete probability function 
and is denoted by 
$p(y|\Do(X=x))$. The mathematical definition of $p(y|\Do(X=x))$ can be
given in terms of a so-called truncated Markov factorization or maybe more
intuitively, by direct plug-in of the intervention value $x$ for variable
$X$ and propagating this intervention value $x$ to all other random
variables including $Y$ in the structural equation model (\ref{SEM0});
precise definitions are given in e.g. \citet{pearl00} or \citet{sgs00}. The
underlying important assumption in the definition of $p(y|\Do(X=x))$ is that the functional forms and error distributions of the structural equations for all the variables $X_j$ which are different from $X$ do not change when making an intervention at $X$. 

A very powerful representation of the intervention distribution is given by
the well-known backdoor adjustment formula.\footnote{For a simple version
  of the formula, skip the text until the second line after formula
  (\ref{backdoor1}).}  
We say that a path in a DAG
$D$ is blocked by a set of nodes $S$ if and only if it contains a chain
$.. \rightarrow m \rightarrow ..$ or a fork $.. \leftarrow m \rightarrow
..$ with $m \in S$ or a collider $.. \rightarrow m \leftarrow ..$ such that
$m \not\in S$ and no descendant of $m$ is in $S$. Furthermore, a set of
variables $S$ is said to satisfy the backdoor criterion relative to
$(X,Y)$ if no node in $S$ is a descendant of $X$ and if $S$ blocks every
path between $X$ and $Y$ with an arrow pointing into $X$. For a set $S$
that satisfies the backdoor criterion relative to $(X,Y)$, the backdoor
adjustment formula reads:  
\begin{equation} \label{backdoor1}
	p(y| \Do(X=x)) = \int p(y| X=x,X_{S}) dP(X_{S}),
\end{equation}
where $p(\cdot)$ and $P(\cdot)$ are generic notations for the density or
distribution \citep[Theorem~3.3.2]{pearl00}. An important special case of
the backdoor adjustment formula is obtained when considering the
adjustment set $S = \pa(j_X)$: if $j_Y \notin \pa(j_X)$, that is, if $Y$ is not in
the parental set of the variable $X$, then: 
\begin{align}\label{backdoor0}
p(y| \Do(X=x)) = \int p(y| X=x,X_{\pa(j_X)}) dP(X_{\pa(j_X)}).
\end{align}
 Thus, if the parental set $\pa(j_X)$ is known, the intervention distribution can be calculated from the standard observational conditional and marginal distributions. 
Our main focus is the expectation
of $Y$ when doing the intervention $\Do(X=x)$, the so-called total effect:
\begin{eqnarray*}
\EE[Y|\Do(X=x)] = \int y \, p(y|\Do(X=x)) dy.
\end{eqnarray*}

A general and often used route for inferring $\EE[Y|\Do(X=x)]$ is as
follows: the directed acyclic graph (DAG) corresponding to the structural
equation model (SEM) is either known or (its Markov equivalence class)
estimated from data; building on this, 
one can estimate the functions in the SEM (edge functions in the DAG), the
error distributions in the SEM, and finally extract an estimate of
$\EE[Y|\Do(X=x)]$ (or bounds of this quantity if the DAG is not
identifiable) from the observational distribution. See for example
\citet{sgs00,pearl00,makapb09,spirtes2010introduction}.

\subsection{Our contribution}\label{subsec.ourcontr}

The new results from this paper should be explained for two different
scenarios and application areas: one where the structure of the DAG $D$ in
the SEM is known, and the other where the structure and the
DAG $D$ are unknown and estimated from data. Of course, the second setting
is linked to the first by treating the estimated as the true known
structure. However, due to estimation errors, a separate discussion is in
place. 

\subsubsection{Structural equation models with known
  structure}\label{subsec.SEMknownstr} 

We consider a general SEM as in (\ref{SEM0}) with known structure in form
of a DAG $D$ but unknown functions $f_j$ and unknown error distributions
for $\eps_j$. As already mentioned before, our focus is on inferring the
total effect
\begin{eqnarray}\label{toteff}
\EE[Y|\Do(X=x)] = \int y \, p(y| \Do(X=x)) dy,
\end{eqnarray}
where $p(y| \Do(X=x))$ is the interventional density (or discrete
probability function) of $Y$ as loosely described in Section~\ref{subsec.basic}. 

The first approach to infer the total effect in (\ref{toteff}) is to
estimate the functions $f_j$ and error distributions for $\eps_j$ in the SEM. It
is then possible to calculate $\EE[Y| \Do(X=x)]$, typically using 
a
path-based method based on the DAG $D$ (see also Section~\ref{subsec.fullpath}). This route is essentially impossible without
putting further assumptions on the functional form of $f_j$ in the SEM
(\ref{SEM0}). For example, one often makes the assumption of additive
errors, and if the cardinality of the parental set $|\pa(j)|$ is large,
additional constraints like additivity of a nonparametric function are
in place to avoid the curse of dimensionality. Thus, by keeping the general
possibly non-additive structure of the functions $f_j$ in the SEM, we have
to reject this approach. 

The second approach for inferring the total effect in (\ref{toteff}) relies
on the powerful backdoor adjustment formula in (\ref{backdoor1}). At first
sight, the problem seems ill-posed because of the appearance of
$p(Y|X=x,X_S)$ for a set $S$ with possibly large cardinality $|S|$. But
since we integrate over the variables $X_S$ in (\ref{backdoor1}), we are
\emph{not} entering the curse of dimensionality. This simple observation is
a key idea of this paper.  
We present an estimation technique for $\EE[Y| \Do(X=x)]$, or other
functionals of $p(y| \Do(X=x))$, using marginal integration which
has been proposed and analyzed for additive regression modeling
\citep{linton1995kernel}. The idea of our marginal integration approach is to first
estimate a fully nonparametric regression of $Y$ versus $X$ and the
variables $X_S$ from a valid adjustment set satisfying the backdoor
criterion (for example the parents of $X$ or a superset
thereof) and then 
average the obtained estimate over the variables $X_S$. We call the
procedure {``\emph{S-mint}''} standing for \emph{m}arginal \emph{int}egration
with adjustment set $S$.   
Our main result in Theorem~\ref{th1} establishes that $\EE[Y|\Do(X=x)]$ can
be inferred via marginal integration with the same rate of convergence as
for one-dimensional nonparametric function estimation for a very large
class of  structural equation models with potentially non-additive
functional forms in the equations. We thereby achieve a major
robustness result against model misspecification, as we only assume some
standard smoothness assumptions but no further conditions on the functional
form or nonlinearity of the functions $f_j$ in the SEM, not even 
additive errors. 
Our main result (Theorem~\ref{th1}) also applies using a 
superset of the true underlying DAG $D$ (i.e. there might be additional
directed edges in the superset), see Section~\ref{subsec.supset}. For
example, such a superset could arise from knowing the order of the
variables (e.g. in a time series context), or an
approximate superset might be available from estimation of the DAG where one
would not care too much about slight or moderate overfitting. 

Inferring $\EE[Y|\Do(X=x)]$ under model-misspecification is the theme of
double robustness in causal inference, typically with a binary treatment
variable $X$ \citep[cf.][]{van2003unified}. There,
misspecification of either the regression or the propensity score
model\footnote{Definitions can be found in
  Section~\ref{subsec.doublerobust}} is allowed but at least one of them
has to be correct to allow for consistent estimation: the terminology  
``double robustness'' is intended to reflect this kind of robustness. In
contrast to double robustness, we achieve here ``full robustness'' where
essentially any form of ``misspecification'' is allowed, in the sense that
 \emph{S-mint} does not require any specification of the
functional form of the structural equations in the SEM. More details are
given in Section~\ref{subsec.doublerobust}.  

\smallskip\noindent
\emph{The local nature of parental sets.}
Our \emph{S-mint} procedure requires the specification of a valid adjustment set $S$: as described in (\ref{backdoor0}), we can always use the parental set
$\pa(j_X)$ if $j_Y \notin \pa(j_X)$. The parental variables are often an
interesting choice for an adjustment set which corresponds to a
\emph{local} 
operation. 
Furthermore, as discussed below, the local nature of the parental sets can
be very beneficial in presence of only approximate knowledge of the true
underlying DAG $D$.  
 
\subsubsection{Structural equation models with unknown structure} 

Consider the SEM (\ref{SEM0}), but now we assume that the DAG $D$ is
unknown.  
For this setting, we propose a two-stage scheme (``\emph{est S-mint}'',
Section~\ref{subsec.twostage}). First, we estimate the structure of the DAG
(or the Markov equivalence class of DAGs) or the order of the
variables from observational data. To do this, all of the current
approaches make further assumptions for the SEM in (\ref{SEM0}), 
see for example
\citet{chick02,teykol05,shim06,kabu07,schmidt07,hoy09,shojaie2010penalized,pbjoja13}. 

We can then infer $\EE[Y|\Do(X=x)]$ as before with \emph{S-mint} model fitting,
but based on an estimated (instead of the true) adjustment set $S$. This
seems often more advisable than using the estimated functions in the SEM,
which are readily available from structure estimation, and pursuing a path-based method with the estimated DAG. Since estimation of (the Markov
equivalence class of) the DAG or of the order of 
variables is often very difficult and with limited accuracy for finite
sample  size, the second stage with \emph{S-mint} model fitting 
is
fairly robust with respect to errors in order- or structure-estimation and model
misspecification, as suggested by our empirical results in
Section~\ref{ssec:CEmodifDAG}. Therefore, such a 
two-stage procedure with structure- or order-search\footnote{We do not make
  use of e.g. estimated edge functions, even if they were implicitly
  estimated for structure-search, as e.g. in \cite{chick02}.} and
subsequent marginal integration leads to reasonably accurate and sometimes
better results. For example, Section~\ref{sec.empirical.add} reports a
comparable performance to the direct CAM method  
\citep{pbjoja13} with subsequent path-based estimation of causal effects, which is based on, or assuming, a correctly specified
additive SEM.\footnote{For a short description of the CAM method, see
  the last paragraph of Section~\ref{subsec.DAGinfer}.} Thus, even if the
\emph{est S-mint} approach with fully 
nonparametric \emph{S-mint} modeling in the second stage is not exploiting the
additional structural assumption of an additive SEM, it exhibits a competitive
performance.

As mentioned in the previous subsection, the
parental sets (or supersets 
thereof) with their local nature are often a very good choice in presence
of estimation errors with respect to inferring the true DAG (or equivalence
class thereof): instead of assuming 
high accuracy for recovering the entire (equivalence class of the) DAG, we
only need to have a reasonably accurate estimate of the much smaller and local
parental set. 

\smallskip\noindent
\emph{A combined structured (or parametric) and fully nonparametric
  approach.} The two-stage \emph{est S-mint} procedure is typically a
combination of a structured nonparametric or parametric approach for
estimating the DAG (or the equivalence class thereof) and the fully
nonparametric \emph{S-mint} method in the subsequent second stage. As
outlined above, it exhibits comparatively good performance. One could think
of pursuing the first stage in a fully nonparametric fashion as well, for
example by using the PC-algorithm with nonparametric conditional
independence tests \citep{sgs00}, see also \citet{song2013kernel}. For
finite amount of data and a fairly large number of variables, this is a
very ambitious if not ill-posed task. In view of this, we almost have to
make additional structural or parametric assumptions for structure learning
of the DAG (or its equivalence class). However, since the fully
nonparametric \emph{S-mint} procedure in the second stage is less sensitive
to incorrect specification of the DAG (or its equivalence class), the
combined approach exhibits better robustness. Vice-versa, if the structural
or parametric model is correct which is used for structural learning in the
first stage, we do not lose much efficiency when ``throwing away'' (or not
exploiting) such structural information in the second stage with
\emph{S-mint}. We only have empirical results to support such accuracy
statements.

\subsection{The scope of possible applications}\label{subsec.applic}

Genetic network inference is a prominent example where causal inference
methods are used; mainly for estimating an underlying network in terms of a
directed graph \citep[cf.][]{smith2002evaluating,husmeier2003sensitivity,friedman2004inferring,yu2004advances}. The goal is very ambitious, namely to recover relevant edges in a complex network
from observational or a few interventional data. This paper does not address this issue: instead of recovering a network (structure), inferring \emph{total} causal or
intervention effects from observational data is a different, maybe more realistic but still very challenging goal in its full generality. Yet making 
progress can be very useful in many areas of applications, notably for prioritizing and
designing future randomized experiments which have a large total effect on a response variable of interest, ranging from molecular biology and bioinformatics (Editorial Nat. Methods, 2010\nocite{editnatmethods10}) to many other fields
including economy, medicine or social sciences. Such model-driven prioritization for 
gene intervention experiments in molecular biology has been experimentally validated with some success \citep{mapb10,steketal12}.

We will discuss an application from molecular biology on a rather
``toy-like'' level in Section~\ref{sec.realdata}. Despite all
simplifying considerations, however, we believe that it indicates a broader
scope of possible applications. When having approximate knowledge of the
parental set of the variables in a potentially large-scale system, one would not
need to worry much about the underlying form of the dependences of (or
structural equations linking) the variables: for quantifying the effect of
single variable interventions, the proposed \emph{S-mint} marginal integration estimator converges with the univariate rate, as stated in (the main result) Theorem~\ref{th1}. 

Quantifying single variable interventions from observational data is indeed
a useful first step.
Further work is needed to address the following issues:
(i) inference in settings with additional
hidden, unobserved variables
\citep[cf.][]{sgs00,zhang2008completeness,shpitser2011efficient,colombetal12};
(ii) inference 
based on both observational and interventional data
\citep[cf.][]{hegeng08,hauser2012characterization,hauser2013two,hauser2013jointly}; and finally (iii) developing sound tools and methods towards more
confirmatory conclusions.  
The appropriate modifications and further developments of our new results
(mainly Theorem~\ref{th1}) towards these points (i)-(iii) are not
straightforward.

\section{Causal effects for general nonlinear systems via backdoor
  adjustment: marginal integration suffices}\label{sec.backdoorgen}

We present here the, maybe surprising, result that marginal integration
allows us to infer the causal effect of a single variable intervention with
a convergence rate as for one-dimensional nonparametric function estimation
in essentially \emph{any} nonlinear structural equation model.  

We assume a structural
equation model (as already introduced in Section~\ref{subsec.basic}) 
\begin{align}\label{SEM}
X_j \leftarrow f_{j}^0(X_{\pa(j)},\eps_j),\ j=1,\ldots ,p,
\end{align}
where $\eps_1,\ldots ,\eps_p$ are independent noise (or innovation) variables,
$\pa(j)$ denotes the set of parents of node $j$ in the underlying DAG $D^0$,
and $f_j^0(\cdot)$ are real-valued (measurable) functions. We emphasize the
true underlying quantities with a superscript ``$^0$''. We assume in this
section that the DAG $D^0$, or at least a (super-) DAG
$D_{\mathrm{super}}^0$ which contains $D^0$ (see
Section~\ref{subsec.supset}), is known. As mentioned earlier, our goal is
to give a 
representation of the expected value of the intervention distribution
$\EE[Y|\Do(X = x)]$ for 
two variables $Y,X \in \{X_1,\ldots ,X_p\}$. 
That is, we want to study the total effect that an intervention at $X$ has on a target variable $Y$.
Let $S$ be a set of variables
satisfying the backdoor criterion relative to $(X,Y)$, 
implying that
\begin{eqnarray*}
p(y| \Do(X=x)) = \int p(y| X=x,X_S) dP(X_S),
\end{eqnarray*}
where $p(\cdot)$ and $P(\cdot)$ are generic notations for the density or
distribution (see Section~\ref{subsec.basic}). Assuming that  
we can interchange the order of integration (cf. part~6 of Assumption~\ref{assu1}) we obtain
\begin{align}\label{backdoor-exp}
\EE[Y| \Do(X=x)] = \int \EE[Y| X=x,X_S] dP(X_S).
\end{align}
This is a function depending on the one-dimensional variable $x$ only and
therefore, intuitively, its estimation should not be much exposed to the curse of
dimensionality. We will argue below that this is indeed the case. 

\subsection{Marginal integration} \label{sec:margint}

Marginal integration is an estimation method which has been primarily
designed for additive and structured regression fitting
\citep{linton1995kernel}. Without any 
modifications though, it is also suitable for the  estimation of $\EE[Y|\Do(X=x)]$ in
(\ref{backdoor-exp}).   

Let $S$ be a set of variables satisfying the backdoor criterion relative to
$(X,Y)$ (see Section~\ref{subsec.basic}) and denote by $s$ the cardinality
of $S$. We use a nonparametric partial local estimator of the multivariate regression
function $m(x,x_S) = \EE[Y| X=x,X_S = x_S]$ 
of the form
\begin{equation} \label{eq:min}
(\hat{\alpha}, \hat{\beta}) = \argminn\limits_{\alpha, \beta}
\sum\limits_{i=1}^n (Y_i - \alpha - \beta (X^{(i)}-x))^2  K_{h_1}(X^{(i)} -x) L_{h_2}(X_S^{(i)}- x_S), 
\end{equation}
where $\hat{\alpha} = \hat{\alpha}(x,x_S), \hat{\beta} = \hat{\beta}(x,x_S)$, $K$ and $L$ are two kernel functions and $h_1, h_2$ the respective
bandwidths, i.e.,
$$
	K_{h_1}(t) = \frac{1}{h_1} K \left( \frac{t}{h_1} \right), \qquad L_{h_2}(t) = \frac{1}{h_2^{s}} L \left( \frac{t}{h_2} \right).
$$  
We obtain the partial local linear estimator at $(x,x_S)$ as $\hat{m}(x,x_S) = \hat{\alpha}(x,x_S)$. 
We then \emph{integrate} over the variables $X_S$ with the
empirical mean and obtain:
\begin{eqnarray}\label{margint}
\hat{\EE}[Y|\Do(X=x)] &=& n^{-1} \sum_{k=1}^n \hat{m}(x,X_S^{(k)}) 
\end{eqnarray}
This is a locally weighted average, with localization through the
one-dimensional variable $x$. For our main theoretical result to hold, we make the following assumptions:
\begin{assu} \label{assu1}
\hfill
\begin{enumerate}
	\item The variables $X_S$ have a bounded support $\text{supp}(X_S)$.
	\item The regression function $m(u, u_S)=\EE[Y| X=u,X_S = u_S]$
          exists and has
          bounded partial derivatives up to order 2 with respect to $u$ and
          up to order $d$ with respect to $u_S$ for $u$ in a neighborhood
          of $x$ and $u_S \in \text{supp}(X_S)$. 
	\item The variables $X,X_S$ have a density $p(.,.)$ with
            respect to Lebesgue measure and 
$p(u, u_S)$ has bounded partial derivatives up to order 2 with respect to $u$ and up to order $d$ with respect to $u_S$. In addition, it holds that 
	$$
		\inf\limits_{\stackrel{u \in x \pm \delta}{x_S \in \text{supp}(X_S)}} p(u,x_S) > 0 \ \ \text{for some } \delta > 0 .
	$$
	\item The kernel functions $K$ and $L$ are symmetric with bounded supports and $L$ is an order-$d$ kernel. 
	\item For $\eps = Y - \EE[Y | X, X_S]$, it holds that $\EE[\eps^4]$ is finite and $\EE[\eps^2 | X=x, X_S = x_S]$ is continuous. Furthermore, for a $\delta > 0$, $\EE[|\eps|^{2 + \delta} \mid X=u]$ is bounded for $u$ in a neighborhood of $x$.
	\item There exists $c < \infty$ such that $\EE[|Y| \vert X=x, X_S = x_S] \leq c$ for all $x_S$.
\end{enumerate}
\end{assu}
Note that part~6 of Assumption~\ref{assu1} is only needed for interchanging the order of integration in (\ref{backdoor-exp}). Due to the bounded support of the variables $X_S$ it is not overly restrictive.

As a consequence, the following result from
\citet{fan1998direct} establishes a 
convergence rate for the estimator as for one-dimensional nonparametric
function estimation. 
\begin{theo}\label{th1}
Suppose that Assumption~\ref{assu1} holds for a set $S$ satisfying the
backdoor criterion relative to $(X,Y)$ in the DAG $D^0$ from model
(\ref{SEM}).
Consider the estimator in (\ref{margint}). Assume that the bandwidths are
chosen such that $h_1, h_2 \rightarrow 0$ with $n h_1 h_2^{2s}/\log^2(n)
\rightarrow \infty$, $h_2^d/h_1^2 \rightarrow 0$, and in addition
satisfying $n h_1 h_2^{s}/\log(n) \rightarrow \infty$ and $h_1^4
\log(n)/h_2^s \rightarrow 0$ (and all these conditions hold when choosing
the bandwidths in a properly chosen optimal range, see below). Then,
\begin{eqnarray*}
\hat{\EE}[Y|\Do(X=x)] - \EE[Y|\Do(X=x)] = O(h_1^2) + O_P(1/\sqrt{nh_1}).
\end{eqnarray*}
\end{theo} 

\begin{prf} The statement follows from \citet[Theorem~1 and
  Remark~3]{fan1998direct}.
\end{prf}

\smallskip
When assuming the smoothness condition $d > s$ for $m(u,u_S)$ with
  respect to the variable $u_S$,  and when choosing $h_1 \asymp n^{-1/5}$
  and $h_2 \asymp n^{-\alpha}$ with $ 2/(5d) < 
\alpha < 2/(5s)$ (which requires $d>s$), all the conditions for the
bandwidths are satisfied: Theorem~\ref{th1} then establishes the
convergence rate $O(n^{-2/5})$ which matches the optimal rate
  for estimation of 
one-dimensional smooth functions having second derivatives, and such a
smoothness condition is
assumed for $m(u,u_S)$ with respect to the variable $u$ in part 2 of Assumption~\ref{assu1}. Thus, the implication is the
important robustness fact that for 
\emph{any} potentially nonlinear structural equation model satisfying the
regularity conditions in Theorem~\ref{th1}, we can estimate the expected
value of the intervention distribution with the same accuracy as in
nonparametric estimation of a smooth function with one-dimensional argument. 
We note, as mentioned already in Section~\ref{subsec.SEMknownstr}, that it would
be essentially impossible to estimate the functions $f_j$ in (\ref{SEM0}) 
in full generality: interestingly, when focusing on inferring the total effect
$\EE[Y| \Do(X=x)]$, the problem is much better posed as demonstrated with
our concrete \emph{S-mint} procedure. 
Furthermore, with the (valid) choice $S = \pa(j_X)$ or an (estimated)
superset thereof, one obtains a procedure that is only based on local
information in the graph: this turns out to be advantageous, see also
Section~\ref{subsec.SEMknownstr}, particularly when the
underlying DAG structure is not correctly specified (see Section~\ref{ssec:CEmodifDAG}).  
We will report about the performance of such an \emph{S-mint} estimation
method in Sections~\ref{sec.empirical.nonadd}~and~\ref{sec.empirical.add}. 
Note that the rate of Theorem~\ref{th1} remains valid (for a slightly modified estimator) if we allow for discrete variables in the parental set of $X$ \citep{fan1998direct}. 

It is also worthwhile to point out that \emph{S-mint} becomes more challenging for inferring multiple variable interventions such
as $\EE[Y|\Do(X_1=x_1,\ X_2=x_2)]$: the convergence rate is then of the
order $n^{-1/3}$ for a twice differentiable regression function. 

\begin{rem}\label{rem2}
Theorem~\ref{th1} generalizes to real-valued transformations $t(\cdot)$ of
$Y$.  By using the argument as in (\ref{backdoor-exp}) and replacing part~6 of Assumption~\ref{assu1} by the corresponding statement for $t(Y)$, we obtain 
\begin{align*}
	\EE[t(Y)|\Do(X=x)] &= \int t(y)p(y|\Do(X=x)) dy  \\
	&= \int \EE[t(Y)|X=x,X_S] dP(X_S).
\end{align*}
For example, for
$t(y) = y^2$ we obtain second moments and we can then estimate the variance
$\mathrm{Var}(Y|\Do(X=x)) = \EE[Y^2| \Do(X=x)] - (\EE[Y| \Do(X=x)])^2$. Or with the
indicator function $t(y) = I(y \le c)$ ($c \in \R$) we obtain a procedure
for estimating $\PP[Y \le c|\Do(X=x)]$ with the same convergence rate as for one-dimensional nonparametric function estimation
using marginal integration of $t(Y)$ versus $X, X_S$.
\end{rem}

\subsubsection{Binary treatment and connection to double
  robustness}\label{subsec.doublerobust}

For the special but important case with binary treatment, where $X \in
\{0,1\}$ and $X_S \in \R^s$ is continuous, we can use marginal integration
as well. We can estimate the regression function $m(x,x_S)$ for $x \in
\{0,1\}$ by using a kernel estimator based on data with the
observed $X^{(k)}=0$ and $X^{(k)}=1$, respectively, denoted by
$\hat{m}(x,x_S)\ (x \in \{0,1\})$. We then integrate over $x_S$ with the
empirical mean $n^{-1} \sum_{k=1}^n \hat{m}(x,X_S^{(k)})\ (x \in
\{0,1\})$. When choosing the bandwidth $h_2$ (for smoothing over
the $X_S$ variables) smaller than for the non-integrated quantity
$m(x,x_S)$, and assuming smoothness conditions, we anticipate the $n^{-1/2}$
convergence rate for estimating $\EE[Y|\Do(X=x)]$ with $x \in \{0,1\}$; see
for example \citet{hall1987estimation} in the context of nonparametric
squared density estimation. We note that this establishes only the optimal
parametric convergence rate but does not generally lead to asymptotic
efficiency. For the case of binary treatment, semiparametric minimax rates
have been established in \citet{robins2009} and asymptotically efficient
methods can be constructed using higher order influence functions \citep{li2011}
or targeted maximum likelihood estimation \citep{vanderLaan11} which both might
be more suitable than marginal integration. 

Theorem~\ref{th1} establishes that \emph{S-mint} is ``fully robust'' against model-misspeci- fication for inferring $\EE[Y|\Do(X=x)]$ or related
quantities as mentioned in Remark~\ref{rem2}. The existing framework of
double robustness is related to the issue of misspecification and we
clarify here the connection.  
One specifies regression models for $\EE[Y|X,X_S] = m(X,X_S)$ for both $X=0$ and $X=1$
and a propensity score \citep{rosenrub83} or inverse probability weighting model
(IPW; \citet{robetal94}): for a binary intervention variable where $X$
encodes ``exposure'' ($X=1$) and ``control'' ($X=0$), the latter is a (often
parametric logistic) model for $\PP[X=1|X_S]$.  
A double robust (DR) estimator for $\EE[Y|\Do(X=x)]$ requires
that either the regression model or the propensity score model is correctly
specified. If both of them are misspecified, the DR estimator is
inconsistent. Double robustness of the augmented IPW approach has been
proved by \citet{scharfetal99} and double robustness in general was further
developed by many others, see e.g. \citet{bangrob05}. The targeted maximum
likelihood estimation (TMLE) framework \citep{vanderLaan11} is also double
robust. It uses a second step where the initial estimate is
modified in order to make it less biased for the target parameter
(e.g. the average causal effect between ``exposure'' and ``control''). If
both, the initial estimator and the treatment mechanism, are consistently
estimated, TMLE can be shown to be asymptotically efficient. TMLE with a
super-learner or also the approach of higher order influence functions
\citep{li2011} can deal with a nonparametric model.  
\citet{robins2009} prove that $s = \mathrm{dim}(X_S) \le
2(\beta_{\mathrm{regr}} + \beta_{\mathrm{propens}})$, where $\beta_{\mathrm{`name'}}$ denotes the smoothness of the regression or
propensity score function, is a
necessary condition for an estimator to achieve the $1/\sqrt{n}$
convergence rate. 

Our \emph{S-mint} procedure is related to these nonparametric approaches:
it differs though in that it deals with a continuous treatment
variable. Similar to the smoothness requirement above we have discussed after 
Theorem \ref{th1} that we can achieve the $n^{-2/5}$ nonparametric optimal
rate (when assuming bounded derivatives up to order 2 of the regression
function with respect to the treatment variable) if $s = \mathrm{dim}(X_S) <
d$, where $d$ plays the role of $\beta_{\mathrm{regr}}$. The condition $s <
d$ is stronger than for the optimal $1/\sqrt{n}$ convergence rate with
binary treatment: however, this could be relaxed to the regime $s < 2d$
when invoking \citet[Remark 1]{fan1998direct}. Therefore, rate
optimal estimation with continuous 
treatment can be achieved under a  ``comparable'' smoothness assumption
as in the binary treatment case.

\subsection{Implementation of marginal integration}\label{subsec.implem}

Theorem~\ref{th1} justifies marginal integration as in (\ref{margint})
asymptotically. One issue is the choice of the two bandwidths $h_1$ and
$h_2$: we cannot rely on cross-validation because $\EE[Y| \Do(X=x)]$ is not
a regression function and is not linked to prediction of a new observation
$Y_{\mathrm{new}}$, nor can we use penalized likelihood techniques
with e.g. BIC since $\EE[Y| \Do(X=x)]$ does not appear in the
likelihood. Besides the difficulty of choosing the smoothing parameters, we
think that addressing such a smoothing problem will become easier, at least
in practice, using 
an iterative boosting approach
 \citep[cf.][]{fried01,pbyu03}.

We propose here a scheme, without complicated tuning of parameters, which
we found to be most stable and accurate in extensive simulations. The idea is to
elaborate on the estimation of the function  
$m(x,x_S) = \EE[Y| X=x,X_{S} = x_S]$, from a simple starting point to more
complex estimates, while the integration over the variables $X_S$ is done
with the empirical mean as in (\ref{margint}).  

We start with the following simple but useful result. 
\begin{prop} \label{propBackdoor}
If $\pa(j_X) = \emptyset$ or if there are no backdoor paths from $j_X$
to $j_Y$ in the true DAG $D^0$ from model (\ref{SEM}), then
\begin{eqnarray*}
\EE[Y|\Do(X=x)] = \EE[Y|X=x].
\end{eqnarray*}
\end{prop}

\begin{prf}
	If there are no backdoor paths from $j_X$ to $j_Y$, the empty set $S=\emptyset$ satisfies the backdoor criterion relative to $(X,Y)$. The statement then directly follows from the backdoor adjustment formula \eqref{backdoor1}.
\end{prf}

\smallskip
We learn from Proposition~\ref{propBackdoor} that in simple cases, a
standard one-dimensional regression estimator for $\EE[Y|X=x]$ would
suffice. On the other hand, we know from the backdoor adjustment formula in
(\ref{backdoor-exp}), that we should adjust with the variables
$X_S$. Therefore, it seems natural to use an \emph{additive} regression
approximation for $m(x,x_S)$ as a simple starting point. If the assumptions
of Proposition~\ref{propBackdoor} hold, such an additive model fit would
yield a consistent estimate for the component of the variable $x$: in fact,
it is asymptotically as efficient as when using one-dimensional function
estimation for $\EE[Y|X=x]$ \citep{horetal06}. If the assumptions of
Proposition~\ref{propBackdoor} would not hold, we can still view an
additive model fit $\hat{m}_{\mathrm{add}}(x,x_S) = \hat{\mu} +
\hat{m}_{\mathrm{add},j_X}(x) + \sum_{j \in S}
\hat{m}_{\mathrm{add},j}(x_j)$ as one of the simplest starting points to
approximate the more complex function $m(x,x_S)$. When integrating out with
the empirical mean as in \eqref{margint}, we obtain the estimate
$\hat{\EE}_{\mathrm{add}}[Y|\Do(X=x)] = \hat{\mu} + \hat{m}_{\mathrm{add},j_X}(x)$. As
motivated above and backed up by simulations,
$\hat{\mu} + \hat{m}_{\mathrm{add},j_X}(x)$ is quite often already a 
reasonable estimator for $\EE[Y|\Do(X=x)]$. 

In the presence of strong interactions between the variables, the additive
approximation may drastically fail though. Thus, we want to implement marginal
integration as follows: starting from $\hat{m}_{\mathrm{add}}$, we
implement $L_2$-Boosting with a nonparametric kernel estimator 
similar to the one in \eqref{eq:min}. More precisely, we compute residuals   
\begin{eqnarray*}
R_1^{(i)} = Y^{(i)} - \hat{m}_{\mathrm{add}}(X^{(i)},X_{S}^{(i)}),\
i=1,\ldots ,n,
\end{eqnarray*}
which, for simplicity, are fitted with a locally constant estimator 
of the form
\begin{equation} \label{eq:locallyconst}
	\hat{\alpha}(x,x_S) = \argminn\limits_{\alpha} \sum\limits_{i=1}^n (R_1^{(i)} - \alpha)^2 K_{h_1}(X^{(i)} - x) L_{h_2}(X_S^{(i)} - x_S).
\end{equation}
The resulting fit is denoted by $\hat{g}_{R_1}(x,x_S) := \hat{\alpha}(x,x_S)$. We add this new
function fit to the previous one and compute again residuals, and we then
iterate the procedure $b_{\mathrm{stop}}$ times. 
To summarize,
for $b=1,\ldots ,b_{\mathrm{stop}}-1$,
\begin{eqnarray*}
& &\hat{m}_1(x,x_S) = \hat{m}_{\mathrm{add}}(x,x_S),\\
& &\hat{m}_{b+1}(x,x_S) = \hat{m}_b(x,x_S) + \hat{g}_{R_b}(x,x_S),\\
& &R_{b+1}^{(i)} = Y^{(i)} - \hat{m}_{b+1}(X^{(i)}, X_S^{(i)}), \quad i=1,...,n.
\end{eqnarray*}
The final estimate for the total causal effect is obtained by marginally
integrating over the variables $X_S$ with the empirical mean as in
\eqref{margint}, that is  
$$
	\hat{\EE}[Y|\Do(X=x)] = n^{-1} \sum\limits_{k=1}^n
        \hat{m}_{b_{\mathrm{stop}}}(x, X_{S}^{(k)}). 
$$
The concrete implementation of the additive model fitting is according to
the default from the \texttt{R}-package \texttt{mgcv}, using penalized thin
plate splines and choosing the regularization parameter in the penalty by
generalized cross-validation, see e.g. \cite{wood06, wood03}. The basis dimension for each smooth is set to $10$.
For the product kernel in \eqref{eq:locallyconst},
we choose $K$ to be a Gaussian 
kernel and $L$ to be a product of Gaussian kernels.  
The bandwidths $h_1$ and $h_2$ in the kernel estimator should be chosen
``large'' to yield an estimator with low variance but typically high
bias. The iterations then reduce the bias. Once we have fixed $h_1$ and
$h_2$ (and this choice is not very important as long as the bandwidths are
``large''), the only regularization parameter is
$b_{\mathrm{stop}}$. It is chosen by the following considerations: for each
iteration we approximate the sum of the differences to the
previous approximation on the set of intervention values $\mathcal{I}$
(typically the nine deciles, see Section \ref{sec.empirical.add}), that is 
\begin{equation}\label{boostdiff}
\sum\limits_{x \in \mathcal{I}} | n^{-1} \sum\limits_{k=1}^n
\hat{g}_{R_b}(x, X_S^{(k)}) | .
\end{equation}
When it becomes reasonably ``small'', and this
needs to be specified depending on the context, we stop the boosting
procedure. 
Such an iterative boosting scheme has the advantage that it is more
insensitive to the choice of $b_{\mathrm{stop}}$ than the original
estimator in \eqref{margint} to the specification of the tuning parameters,
and in addition, boosting adapts to some extent to different smoothness in
different directions (variables). All these ideas are presented at various
places in the boosting literature, particularly in
\citet{fried01,pbyu03,BuhlmannHothorn06}. In Section~\ref{subsec:bdpaths} we provide an example of a DAG with backdoor
paths, where the additive approximation is incorrect and several boosting
iterations are needed to account for interaction effects between the
variables. The implementation of our method is summarized in
Algorithm~\ref{alg0}: we call it also \emph{S-mint}, and we use it for all
our empirical results in Sections
\ref{sec.empirical.nonadd} - \ref{sec.realdata}. 

\begin{algorithm}[!htb]
\begin{algorithmic}[1]
\IF{$S = \emptyset$ is a valid adjustment set (for example, if $\pa(j_X) = \emptyset$)} 
\STATE Fit an additive regression of $Y$ versus $X$ to obtain $\hat{m}_{\text{add}}$
\RETURN $\hat{m}_{\text{add}}$ 
\ELSE
\STATE Fit an additive regression of $Y$ versus $X$ and the adjustment set
variables $X_S$ to obtain $\hat{m}_1 = \hat{m}_{\text{add}}$
\FOR{$b=2,...,b_{\text{stop}}-1$} 
\STATE Apply $L_2$-boosting to capture deviations from an additive
regression model:  
\STATE (i)  \hspace{0.2cm} Compute residuals $R_{b} = Y - \hat{m}_{b}$
\STATE (ii) \hspace{0.14cm} Fit the residuals with the kernel estimator
 \eqref{eq:locallyconst} to obtain $\hat{g}_{R_b}$  
\STATE (iii) \hspace{0.03cm} Set $\hat{m}_{b+1} = \hat{m}_b + \hat{g}_{R_b}$
\ENDFOR
\RETURN Do marginal integration: output $\frac{1}{n} \sum_{k=1}^n
  \hat{m}_{b_{\text{stop}}}(x,X_S^{(k)})$
\ENDIF
\end{algorithmic}
\caption{S-mint}\label{alg0}
\end{algorithm}

We note the following about $L_2$-boosting: if the initial estimator is a
weighted mean $\hat{m}_1(x,x_S) = \sum_{i=1}^n w_i^{(1)}(x,x_S) Y_i $ with
$\sum_{i=1}^n w_i^{(1)}(x,x_S) = 1$ (e.g. many additive function estimators
are of this form), then, since the kernel estimator
$\hat{g}_{R_b}$ in the boosting step 9 is a weighted mean too, $\hat{m}_b(x,x_S) = \sum_{i=1}^n w_i^{(b)}(x,x_S) Y_i$ is a weighted
mean. Thus, $L_2$-boosting has the form of a weighted mean estimator. When
using kernel estimation for $\hat{g}_{R_b}$, the boosting estimator
$\hat{m}_{b_{\mathrm{stop}}}$ is related to an estimator with a higher
order kernel \citep{di2008boosting} which depends on the 
bandwidth in $\hat{g}_{R_b}$ and the number of boosting iterations in a
rather non-explicit way. Establishing the theoretical properties of the
$L_2$-boosting estimator $\hat{\EE}[Y|\Do(X=x)] = n^{-1} 
\sum\limits_{k=1}^n \hat{m}_{b_{\mathrm{stop}}}(x, X_{S}^{(k)})$ is beyond
the scope of this paper. 

\subsection{Knowledge of a superset of the DAG}\label{subsec.supset}

It is known that a superset of the parental set $\pa(j_X)$ suffices
for the backdoor adjustment in (\ref{backdoor0}). To be precise, let 
\begin{eqnarray}\label{supset.ord}
S(j_X) \supseteq \pa(j_X)\ \mbox{with}\ S(j_X) \cap \mathrm{de}(j_X) =
\emptyset,
\end{eqnarray}
 where  $\mathrm{de}(j_X)$ are the descendants of $j_X$ (in the true DAG
$D^0$). For example, $S(j_X)$ could be the parents of $X$ in a superset of the
true underlying DAG (a DAG with additional edges relative to the true
DAG). We
can then choose the adjustment set $S$ in (\ref{margint}) as $S(j_X)$ and
Theorem~\ref{th1} still holds true, assuming that the cardinality $|S(j_X)|
\le M < \infty$ is bounded. Thus, with the choice $S = S(j_X)$, we can use
marginal integration by marginalizing over the variables $X_{S(j_X)}$. 

A prime example where we are provided with a superset $S(j_X)
\supseteq \pa(j_X)$ with $S(j_X) \cap \mathrm{de}(j_X) = \emptyset$ is when
we know the order of the variables and can deduct an approximate superset
of the parents from that. 
When the variables are ordered with $X_j \prec
X_k$ for $j < k$, 
we would use 
\begin{align}\label{supset.ord1}
S(j_X) = \{k; j_X-p_{\max} \leq k< j_X\} \supseteq \pa(j_X),
\end{align}
where ``$\prec$'' and $p_{max}$ denote the order relation among the
variables and an upper bound on the size of the superset to ensure that $S(j_X) \supseteq
\pa(j_X)$. 
\begin{corr}\label{cor1}
Consider the estimator in (\ref{margint}) and assume the conditions of
Theorem~\ref{th1} for the variables $Y, X$ and $X_{S(j_X)}$ with $S(j_X)$ in
(\ref{supset.ord}) or $S(j_X)$ as  in (\ref{supset.ord1}) for ordered
variables.  
Then,  
\begin{eqnarray*}
\hat{\EE}[Y|\Do(X=x)] - \EE[Y|\Do(X=x)] = O(h_1^2) + O_P(1/\sqrt{nh_1}).
\end{eqnarray*}
\end{corr} 

\begin{prf}
	The statement is an immediate consequence of Theorem~\ref{th1}, as
        $S(j_X)$ in \eqref{supset.ord} and \eqref{supset.ord1} satisfies
        the backdoor criterion relative to $(X,Y)$. 
\end{prf}

\section{Path-based methods}\label{sec.CAMpath} 

We assume in the following until Section~\ref{subsec.twostage} that we know the
true DAG and all true functions and error distributions in the general SEM
(\ref{SEM0}). Thus, in contrast to Section~\ref{sec.backdoorgen}, we have
here also knowledge of the entire structure in form of the DAG $D^0$ (and not
only a valid adjustment set $S$ assumed for Theorem~\ref{th1}). This allows us 
to infer  
$\EE[Y|\Do(X=x)]$ in different ways than the generic \emph{S-mint}
regression from Section~\ref{sec.backdoorgen}. The motivation to look at
other methods is driven by potential gains in statistical accuracy when
including the additional information of the functional form or of the
entire DAG in the structural equation model. We will empirically
analyze this issue in Section~\ref{sec.empirical.add}. 

\subsection{Entire path-based method from root nodes}\label{subsec.fullpath}

Based on the true DAG, the variables can always be ordered such that 
\begin{align*}
X_{j_1} \prec X_{j_2} \prec \ldots \prec X_{j_p}.
\end{align*}
Denote by $j_X$ and $j_Y$ the indices of the variables $X$ and
$Y$, respectively. 

If $X$ is not an ancestor of $Y$, we trivially know that $\EE[Y|\Do(X=x)] =
\EE[Y]$. If $X$ is an ancestor of $Y$ it must hold that $j_X < j_Y$. We can
then generate the intervention distribution of the random variables
${X_{j_1} \prec X_{j_2} \prec \ldots \prec Y\,|\,\Do(X=x)}$ in the model (\ref{SEM0}) as follows \citep[Def. 3.2.1.]{pearl00}: 
\begin{description}
\item[Step 1] Generate $\eps_{j_1},\ldots ,\eps_{j_Y}$.
\item[Step 2] Based on Step 1, recursively generate:
\begin{align*}
& X_{j_1} \leftarrow \eps_{j_1},\\
& X_{j_2} \leftarrow f^0_{j_2}(X_{\pa(j_2)},\eps_{j_2}),\\
&\ldots ,\\
& X_{j_X} \leftarrow x,\\
& \ldots,\\
& X_{j_Y} \leftarrow f_{j_Y}^0(X_{\pa(j_Y)},\eps_{j_Y}).
\end{align*}
\end{description}
Instead of an analytic expression for $p(Y|\Do(X=x))$ by integrating out
over the other variables $\{X_{j_k};\ k \neq j_X,j_Y\}$ we rather rely on
simulation. We draw $B$ samples $Y^{(1)} = X_{j_Y}^{(1)},\ldots ,Y^{(B)} =
X_{j_Y}^{(B)}$ by $B$ independent simulations
of Steps~1-2 above and we then approximate, for $B$ large,
\begin{align*}
\EE[Y|\Do(X=x)] \approx B^{-1} \sum_{b=1}^B Y^{(b)}.
\end{align*}
Furthermore, the simulation technique allows to obtain the distribution of
\linebreak $p(Y|\Do(X=x))$ via e.g. density estimation or histogram approximation
based on $Y^{(1)},\ldots ,Y^{(B)}$. 

The method has an implementation in Algorithm~\ref{alg1}
which uses propagation of simulated random variables along directed paths
in the DAG. The method exploits the entire paths in the DAG from the root
nodes to node $j_Y$ corresponding to the random variable $Y$. Figure~\ref{fig:illustAlgorithms2} provides an
  illustration.  
\begin{algorithm}[!htb]
\begin{algorithmic}[1]
\STATE If there is no directed path from $j_X$ to $j_Y$, the interventional
and observational quantities coincide: $p(Y|\Do(X=x))
\equiv p(Y)$ and $\EE[Y|\Do(X=x)] \equiv \EE[Y]$. 
\STATE If there is a directed path from $j_X$ to $j_Y$, proceed with
steps 3-9. 
\STATE Set $X=X_{j_X} = x$ and delete all in-going arcs into $X$. 
\STATE Find all directed paths from root nodes (including $j_X$) to $j_Y$, and denote them by $p_1,\ldots ,p_q$.  
\FOR{$b = 1,\ldots ,B$}
\STATE for every path, recursively simulate the corresponding random variables according to the order of the variables in the DAG: 
\begin{enumerate}
\item[(i)] Simulate the random variables corresponding to the root nodes of $p_1,\ldots
  ,p_{q}$; 
\item[(ii)] Simulate in each path $p_1,\ldots, p_q$ the random variables
  following the root nodes; proceed recursively, according to the order of
  the variables in the DAG. 
\item[(iii)] Continue with the recursive simulation of random variables
  until $Y$ is simulated. 
\end{enumerate}
\STATE Store the simulated variable $Y^{(b)}$.
\ENDFOR
\STATE Use the simulated sample $Y^{(1)},\ldots ,Y^{(B)}$ to approximate
the intervention distribution $p(y|\Do(X=x))$ or its expectation
$\EE[Y|\Do(X=x)]$. 
\end{algorithmic}
\caption{Entire path-based algorithm for simulating the intervention
  distribution}\label{alg1}
\end{algorithm}

When having estimates of the true DAG, all true functions
and error distributions in the additive structural equation model
(\ref{addSEM}), we would use the procedure above based on these estimated
quantities; 
for the error distributions, we either use the
estimated variances in Gaussian distributions, or we rely on bootstrapping
residuals from the structural equation model (typically with residuals
centered around zero).

\subsection{Partially path-based method with short-cuts} 
\label{subsec.partial}

Mainly motivated by computational considerations (see also Section~\ref{subsec.local}), a modification of the
procedure in Algorithm~\ref{alg1} is valid. Instead of considering all
paths from root nodes to $j_Y$ (corresponding to variable $Y$), we only
consider all paths from $j_X$ (corresponding to variable $X$) to $j_Y$ and
simulate the random variables on these paths $p'_1,\ldots
,p'_{m}$. Obviously, in comparison to Algorithm~\ref{alg1}, $m \le q$
and every $p'_k$ corresponds to a path $p_r$ for 
an $r \in \{1,\ldots
,q\}$.  

Every path $p'_k$ is of the form
\begin{align*}
j_X = j_{k,1} \rightarrow j_{k,2} \rightarrow \ldots \rightarrow j_{k,\ell_k-1}
\rightarrow j_{k,\ell_k} = j_Y,
\end{align*}
having length $\ell_k$.
For recursively simulating the random variables on the paths $p'_1,\ldots,
p'_{m}$ we start with setting 
\begin{align*}
X = X_{j_X} \leftarrow x. 
\end{align*}
Then we recursively simulate the random variables corresponding to all the
paths $p'_1,\ldots ,p'_m$  according to the order
of the variables in the DAG. 
For each of these random variables $X_j$ with $j \in \{p'_1,\ldots ,p'_m\}$
and $j \neq j_X$, we
need the corresponding parental variables and error terms in 
\begin{equation*}
X_{j} \leftarrow f^0_{j}(X_{\pa(j)},\eps_j),
\end{equation*}
where for every $k \in \pa(j)$ we set
\begin{align}\label{simulate2}
& X_k = \begin{cases}\mbox{the previously simulated value}, &\mbox{if}\ k \in \{p'_1,\ldots ,p'_m\},\\
\mbox{bootstrap resampled}\ X_k^*,& \mbox{otherwise}, \end{cases}
\end{align}
where the bootstrap resampling is with replacement from the entire
data. The errors are simulated according to the error distribution.  

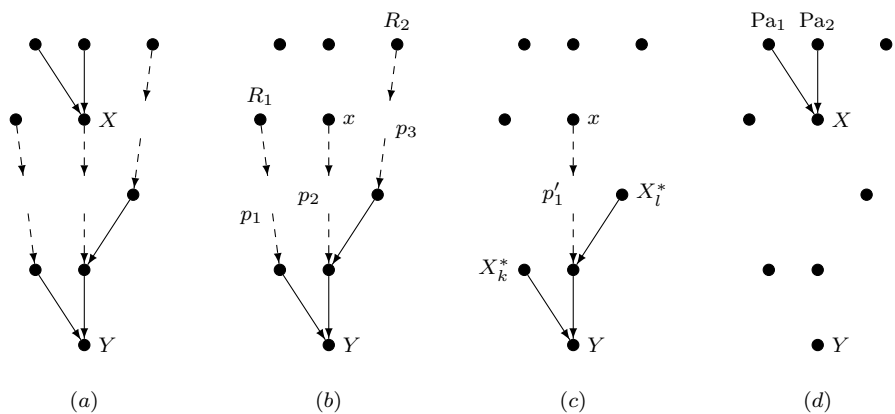
\begin{figure}[h]
\begin{center}

\[\begin{tikzpicture}[x=1.3cm, y=1cm,
    every edge/.style={
        draw,
        postaction={decorate,
                    decoration={markings,mark=at position 1 with {\arrow{latex}}}
                   }
        }
]
	\vertex (ul) at (0,4) {};
	\vertex (um) at (0.5,4) {};
	\vertex (ur) at (1.2,4) {};
	\vertex (ll) at (-0.2,3) {};
	\vertex (lm) at (0.5,3) [label=right:$X$] {};
	\vertex (kr) at (1,2) {};
	\vertex (nl) at (0,1) {};
	\vertex (nr) at (0.5,1) {};
	\vertex (m) at (0.5,0) [label=right:$Y$] {};
	
	\path
		(ul) edge (lm)
		(um) edge (lm)
		(kr) edge (nr)
		(nr) edge (m)
		(nl) edge (m)
	;
	\path [dashed] (lm) edge (0.5,2.25);
	\path [dashed] (0.5,1.75) edge (nr);
	\path [dashed] (ll) edge (-0.125,2.25);
	\path [dashed] (-0.075,1.75) edge (nl);
	\path [dashed] (ur) edge (1.125,3.25);
	\path [dashed] (1.075,2.75) edge (kr);

	\vertex (ul) at (2.5,4) {};
	\vertex (um) at (3,4) {};
	\vertex (ur) at (3.7,4) [label=above:$R_2$] {};
	\vertex (ll) at (2.3,3) [label=above:$R_1$] {};
	\vertex (lm) at (3,3) [label=right:$x$] {};
	\vertex (kr) at (3.5,2) {};
	\vertex (nl) at (2.5,1) {};
	\vertex (nr) at (3,1) {};
	\vertex (m) at (3,0) [label=right:$Y$] {};
	
	\coordinate [label=above:$p_1$] (C) at (2.21,1.5);	
	\coordinate [label=above:$p_2$] (D) at (2.8,1.75);	
	\coordinate [label=above:$p_3$] (E) at (3.8,2.65);	

	\path
		(kr) edge (nr)
		(nr) edge (m)
		(nl) edge (m)
	;
	\path [dashed] (lm) edge (3,2.25);
	\path [dashed] (3,1.75) edge (nr);
	\path [dashed] (ll) edge (2.375,2.25);
	\path [dashed] (2.425,1.75) edge (nl);
	\path [dashed] (ur) edge (3.625,3.25);
	\path [dashed] (3.575,2.75) edge (kr);

	\vertex (ul) at (5,4) {};
	\vertex (um) at (5.5,4) {};
	\vertex (ur) at (6.2,4) {};
	\vertex (ll) at (4.8,3) {};
	\vertex (lm) at (5.5,3) [label=right:$x$] {};
	\vertex (kr) at (6,2) [label=right:$X_l^*$] {};
	\vertex (nl) at (5,1) [label=left:$X_k^*$] {};
	\vertex (nr) at (5.5,1) {};
	\vertex (m) at (5.5,0) [label=right:$Y$] {};
	
	\coordinate [label=above:$p_1'$] (F) at (5.3,1.75);		
	
	\path
		(kr) edge (nr)
		(nr) edge (m)
		(nl) edge (m)
	;
	\path [dashed] (lm) edge (5.5,2.25);
	\path [dashed] (5.5,1.75) edge (nr);	

	\vertex (ul) at (7.5,4) [label=above:$\text{Pa}_1$] {};
	\vertex (um) at (8,4) [label=above:$\text{Pa}_2$] {};
	\vertex (ur) at (8.7,4) {};
	\vertex (ll) at (7.3,3) {};
	\vertex (lm) at (8,3) [label=right:$X$] {};
	\vertex (kr) at (8.5,2) {};
	\vertex (nl) at (7.5,1) {};
	\vertex (nr) at (8,1) {};
	\vertex (m) at (8,0) [label=right:$Y$] {};
		
	\path
		(ul) edge (lm)
		(um) edge (lm)
		;

	\coordinate [label=above:$(a)$] (a) at (0.5,-1);	
	\coordinate [label=above:$(b)$] (b) at (3,-1);	
	\coordinate [label=above:$(c)$] (c) at (5.5,-1);	
	\coordinate [label=above:$(d)$] (d) at (8,-1);	
\end{tikzpicture}\]
\caption{(a) True DAG $D^0$. (b) Illustration of Algorithm 1. $X$ is set to
  $x$, the roots $R_1,R_2$ and all paths from the root nodes and $X$ to $Y$ are
  enumerated (here: $p_1,p_2,p_3$). The interventional distribution at node
  $Y$ is obtained by propagating samples along the three paths. (c)
  Illustration of Algorithm 2. $X$ is set to $x$ and all directed paths
  from $X$ to $Y$ are labeled (here: $p_1'$). In order to obtain the
  interventional distribution at node $Y$, samples are propagated along the
  path $p_1'$ and bootstrap resampled $X_k^*$ and $X_l^*$ are used
  according to \eqref{simulate2}. (d) Illustration of the \emph{S-mint}
  method with adjustment set $S = \pa(j_X)$: it only uses information about $Y, X$
  and the parents of $X$ (here: $\text{Pa}_1, \text{Pa}_2$).}
\label{fig:illustAlgorithms2}
\end{center}
\end{figure}

We summarize the procedure in Algorithm~\ref{alg2} and Figure~\ref{fig:illustAlgorithms2} provides an illustration.  
\begin{algorithm}[!htb]
\begin{algorithmic}[1]
\STATE If there is no directed path from $j_X$ to $j_Y$, the interventional
and observational quantities coincide: $p(Y|\Do(X=x))
\equiv p(Y)$ and $\EE[Y|\Do(X=x)] \equiv \EE[Y]$. 
\STATE If there is a directed path from $j_X$ to $j_Y$, proceed with steps 3-9. 
\STATE Set $X = X_{j_X} = x$ and delete all in-going arcs into $X$.
\STATE Find all directed paths from $j_X$ to $j_Y$, and denote them by
$p'_1,\ldots ,p'_m$.  
\FOR{$b = 1,\ldots ,B$}
\STATE for every path, recursively simulate the corresponding random variables according to the order of the variables in the DAG: 
\begin{enumerate}
\item[(i)] Simulate in each path $p'_1,\ldots, p'_m$ the random variables
  following the node $j_X$; proceed recursively as described in (\ref{simulate2}) according to the order of the variables in the DAG. 
\item[(ii)] Continue with the recursive simulation of random variables
  until $Y$ is simulated.
\end{enumerate}
\STATE Store the simulated variable $Y^{(b)}$.
\ENDFOR
\STATE Use the simulated sample $Y^{(1)},\ldots ,Y^{(B)}$ to approximate
the intervention distribution $p(y|\Do(X=x))$ or its expectation
$\EE[Y|\Do(X=x)]$. 
\end{algorithmic}
\caption{Partially path-based algorithm for simulating the intervention
  distribution}\label{alg2}
\end{algorithm}

\begin{prop}\label{prop1}
Consider the population case where the bootstrap resampling in (\ref{simulate2})
yields the correct distribution of the random variables $X_1,\ldots
,X_p$. Then, as $B \to \infty$, the partially path-based Algorithm~\ref{alg2} yields the correct intervention distribution $p(y|\Do(X=x))$ and
its expected value $\EE[Y|\Do(X=x)]$.
\end{prop}
\begin{prf}
The statement of Proposition~\ref{prop1} directly follows from the definition of the intervention distribution in a structural equation model.
\end{prf}

The same comment as in Section~\ref{subsec.fullpath} applies here: when
having estimates of the quantities in the additive structural equation model
(\ref{addSEM}), we would use Algorithm~\ref{alg2} based on the plugged-in
estimates. The computational benefit of using Algorithm~\ref{alg2} instead of
Algorithm~\ref{alg1} is illustrated in Figure~\ref{fig:CEknownDAG}.

\subsection{Degree of localness}\label{subsec.local}
  
We can classify the different methods according to the degree of which the
entire or only a small (local) fraction of the DAG is used. Algorithm~\ref{alg1} is a rather global procedure as it uses entire paths from root
nodes to $j_Y$. Only when $j_Y$ is close to the relevant root nodes, the
method does involve a smaller aspect of the DAG. Algorithm~\ref{alg2} is of
semi-local nature as it does not require to consider paths going from root
nodes to $j_Y$: it only considers paths from $j_X$ to $j_Y$ and all
parental variables along these paths. The \emph{S-mint} method based on marginal integration described in Section~\ref{sec.backdoorgen} and Theorem~\ref{th1} is of very local character as it only requires the knowledge of $Y, X$ and the parental set $\pa(j_X)$ (or a superset thereof) but no further information about paths from $j_X$ to $j_Y$. 

In the presence of estimation errors, a local method might be more ``reliable''
as only a smaller fraction of the DAG needs to be approximately correct;
global methods, in contrast, require that entire paths in the DAG are
approximately correct. The local versus global issue is illustrated
qualitatively in Figure~\ref{fig:illustAlgorithms2}, and empirical results
about statistical accuracy of the various methods are given in Section~\ref{sec.empirical.add}.

\subsection{Estimation of DAG, edge functions and error distributions}\label{subsec.DAGinfer}

With observational data, in general, it is impossible to infer the true
underlying DAG $D^0$ in the structural equation model (\ref{SEM}), or its
parental sets, even as sample size tends to infinity. One
can only estimate the Markov equivalence class of the true DAG, assuming
faithfulness of the data-generating distribution, see
\citet{sgs00,pearl00,chick02,kabu07,sarpet12,pbcausal13}. The
latter three references focus on the high-dimensional Gaussian scenario with the
number of random variables $p \gg n$ but assuming a sparsity condition in
terms of the maximal degree of the skeleton of the DAG $D^0$. The edge
functions and error variances can then be estimated for every DAG member in
the Markov equivalence class by pursuing regression of a variable versus
its parents. 

However, there are interesting exceptions regarding identifiability of the
DAG from the observational distribution. For nonlinear structural equation
models with additive error terms, it is 
possible to infer the true underlying DAG from infinitely many
observational data \citep{hoy09,petersetal13}. Various methods have been
proposed to infer the true underlying DAG $D^0$ and its corresponding
functions $f_{j}^0(\cdot)$ and error distributions of the $\eps_j$'s:
see for example
\citet{imoto02,hoy09,petersetal13,pbjoja13,vdg13,nowzopb13} (the fourth and
fifth references are considering high-dimensional scenarios). Another
interesting   
class of models where the DAG $D^0$ can be identified 
from the observational data distribution are linear structural equation models
with non-Gaussian noise \citep{shim06}, or with Gaussian noise but equal or
approximately equal error variances \citep{sarpet12,petbu13,lobu13} (the
first and third references are considering the high-dimensional setting). 

As an example of a model with identifiable structure (DAG $D^0$) we can 
specialize (\ref{SEM}) to an additive structural equation model of the form
\begin{align}\label{addSEM}
X_j \leftarrow \sum_{k \in \pa(j)}f^0_{jk}(X_k) + \eps_j,\ j=1,\ldots ,p,
\end{align}
where $\eps_1,\ldots,\eps_p$ are independent with $\eps_j \sim {\cal
  N}(0,(\sigma_j^0)^2)$, and the true 
underlying DAG is denoted by $D^0$. 
This model is used for all numerical comparisons of the \emph{S-mint} procedure and the path-based algorithms in Section~\ref{sec.empirical.add}. 
Estimation of the unknown quantities $D^0$, $f_{jk}^0$ and error variances
$(\sigma_j^0)^2$ can be done with the ``CAM'' method outlined below and
used for the empirical results in 
Section~\ref{ssec:CEestimDAG} in connection with the two-stage procedure \emph{est
  S-mint} that will be introduced in Section~\ref{subsec.twostage}.    

\smallskip\noindent
\emph{The CAM method \citep{pbjoja13}.} The abbreviation ``CAM'' stands for
Causal Additive Model, the additive structural equation model in
(\ref{addSEM}). The CAM method is a nonparametric technique fitting smooth
additive functions and Gaussian error terms in such an additive SEM. The
unknown DAG $D^0$ is estimated by restricted maximum likelihood: the
restriction is on a space of sparse graphs (which can be determined by
e.g. neighborhood selection with Group Lasso for an undirected additive
association graph) and there is no further regularization of such a
restricted MLE. The CAM method is consistent, even in the high-dimensional
scenario with $p \gg n$ but assuming a sparse underlying true DAG.

\subsection{Two-stage procedure: \emph{est S-mint}}\label{subsec.twostage}

If the order of the variables or (a superset of) the parental set is
unknown, we have to estimate it from observational data; 
this leads to the following two-stage procedure described here for the case
where the parental set $\pa(j_X)$ is identifiable:
\begin{description}
\item[Stage 1] Estimate a superset of the parental set $S(j_X)$ (defined in
  \eqref{supset.ord}) from observational data.  
\item[Stage 2] Based on the estimate $\hat{S}(j_X)$, run \emph{S-mint}
  regression with $S = \hat{S}(j_X)$. 
\end{description}
Even if in Stage 1 one would also obtain estimates of functions in 
a specified SEM besides an estimate of $S(j_X)$, we would not use the 
estimated functions in Stage 2. We present empirical results for the
\emph{est S-mint} procedure in connection with the CAM method for
Stage 1 for estimating a valid adjustment set $S(j_X)$ in
Section~\ref{ssec:CEestimDAG}.

If the parental set $\pa(j_X)$ is not 
identifiable (see Section~\ref{subsec.DAGinfer}), one could
apply Stage 1 to obtain a set $\{\hat{S}(j_X)^{(1)},\ldots ,\hat{S}(j_X)^{(c_j)}\}$ such that the parental sets from each Markov-equivalent DAG would be contained in at least one of the $\hat{S}(j_X)^{(k)}$ for some $k$. Stage 2 would then be performed for all estimates $\{\hat{S}(j_X)^{(1)},\ldots
,\hat{S}(j_X)^{(c_j)}\}$ and one could then derive bounds of the quantity
$\EE[Y|\Do(X=x)]$ in the spirit of the approach of \citet{makapb09}. 

In Section~\ref{subsec.summaryemp} we will give some intuition why the two stage
\emph{est S-mint} is often leading to better and more reliable
results than (at least some) other methods which rely on path-based
estimation. 

\section{Empirical results: non-additive structural equation models} \label{sec.empirical.nonadd}
In this section we provide simple proof-of-concept examples for the
generality of the proposed \emph{S-mint} estimation method
(Algorithm~\ref{alg0}). In particular, the robustness of \emph{S-mint} is
experimentally validated for models where the structural equation model is
not additive as in~\eqref{addSEM} but given in its general form~\eqref{SEM}. We make a naive
comparison to path-based methods which are inconsistent
due to incorrect specification of the model in
Section~\ref{subsec:nobdpaths}. However, taking the view of 
classical robustness \citep[cf.][]{hampel2011robust}, we consider a
complementary and interesting issue 
in Section~\ref{sec.empirical.add}: namely the ``efficiency'' of a robust
procedure in comparison to other methods relying on the
correct model specification.

In Section~\ref{subsec:nobdpaths} we empirically show that the path-based
methods based on the wrong additive model assumption in (\ref{addSEM}) may
fail even in the absence of backdoor paths where the \emph{S-mint} method
boils down to estimation of an additive model. In
Section~\ref{subsec:bdpaths} we add backdoor paths to the graph and a
strong interaction term to the corresponding structural equation model. We
then empirically show that \emph{S-mint} manages to approximate the true
causal effect, whereas fitting only an additive regression
fails. Section~\ref{subsec:nonaddnoise} contains an example that
demonstrates a good performance of \emph{S-mint} even in the presence of
non-additive noise in the structural equation model. Finally,
Section~\ref{subsec:bandwidth} empirically illustrates issues with the
fixed choice of the bandwidths in the product kernel in~\eqref{eq:locallyconst} in
some cases. 

\subsection{Causal effects in the absence of backdoor paths} \label{subsec:nobdpaths}
\begin{figure}[!htb]
\begin{minipage}{0.39\textwidth}
\[\begin{tikzpicture}[x=1.3cm, y=1cm,
    every edge/.style={
        draw,
        postaction={decorate,
                    decoration={markings,mark=at position 1 with {\arrow{latex}}}
                   }
        }
        ]
        \tikzstyle{every state}=[fill=white,draw=black,text=black, inner sep=0.25pt, minimum size=15pt]

	\node[state] (X3) {$X_3$};
  \node[state]         (X2) [above right=1.3cm and 0.5cm of X3] {$X_2$};
  \node[state]         (X1) [above left=1.3cm and 0.5cm of X3] {$X_1$};
  \node[state]         (Y) [below=1.3cm of X3] {$Y$};

	\path
		(X1) edge (X3)
		(X2) edge (X3)
		(X3) edge (Y) ;
			
\end{tikzpicture}\]
\end{minipage}
\begin{minipage}{0.59\textwidth}
\begin{center}
\includegraphics[width=0.7\textwidth]{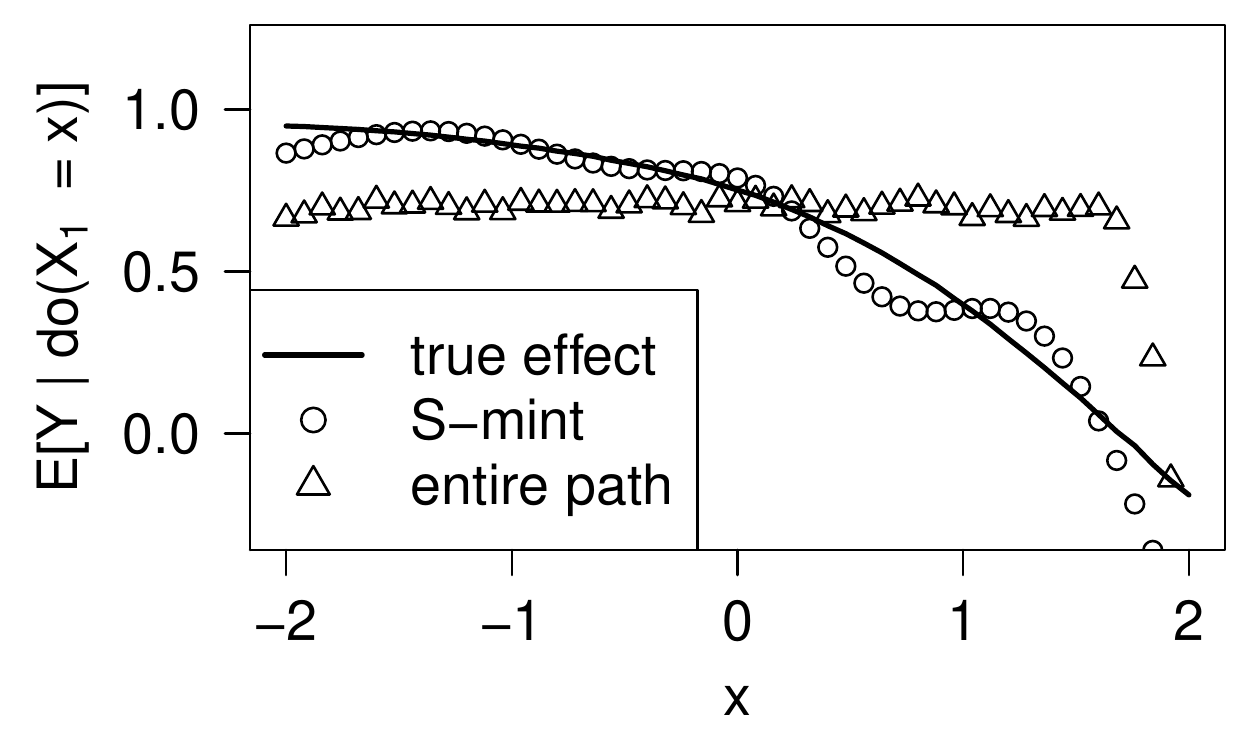} \\
\includegraphics[width=0.7\textwidth]{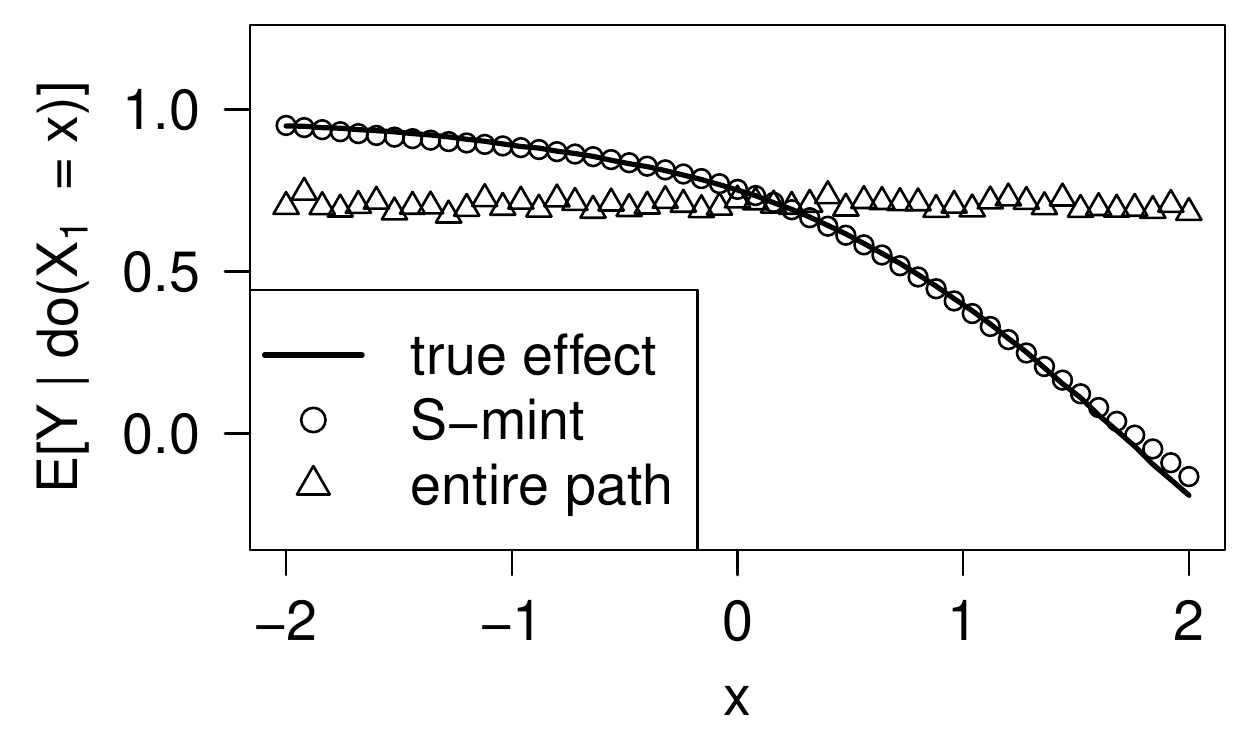} 
\end{center}
\end{minipage}
\caption{Left: DAG corresponding to the structural equation model (\ref{nonadd-SEM}). Right:
  \emph{S-mint} regression estimates of $\EE[Y | \Do(X_1=x)]$ for the model in 
  (\ref{nonadd-SEM}), with $S = S(j_X = 1) = \emptyset$, based on one
  representative sample each for sample sizes 
  $n=500$ (top) and $n=10'000$ 
  (bottom). \emph{S-mint} regression is consistent while the entire path-based
  method with a   misspecified additive SEM (Algorithm~\ref{alg1}) is not. The 
  relative squared errors (over the $51$ points $x$) are $0.013$
  for \emph{S-mint} regression and $6.239$ for the entire path-based method, both for $n=10000$.}
  \label{fig.non-add}
\end{figure}
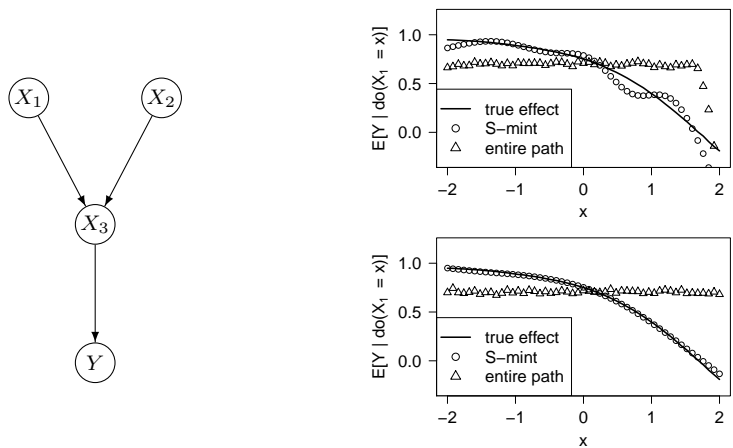

First let us illustrate the sensitivity of the path-based methods with
respect to model specification, using a simple example of a $4$-node graph
with no backdoor paths between $X_1 = X$ and $Y$ (see Figure~\ref{fig.non-add}).
We consider a corresponding (non-additive) structural
equation model of the form 
\begin{align}\label{nonadd-SEM}
	X_1 &\leftarrow \eps_1 \nonumber\\
	X_2 &\leftarrow \eps_2 \nonumber\\
	X_3 &\leftarrow \cos(4\cdot (X_1 + X_2)) \cdot \exp(X_1/2 + X_2/4)
        + \eps_3 \nonumber \\
	Y \ &\leftarrow \cos(X_3)\cdot \exp(X_3/4) + \eps_4
\end{align}
where $\eps_j \sim \mathcal{N}(0, \sigma_j^2)$ with $\sigma_1 = \sigma_2 =
0.7$ and $\sigma_3 = \sigma_4 = 0.2$. We generate $n$ samples from this
model. 
From Proposition~\ref{propBackdoor} we know that for $j \in \{1,2,3\}$,
fitting an additive regression of $Y$ versus $X_j$ and $X_{\pa(j)}$
suffices to obtain the causal effect $\EE[Y|\Do(X_j=x)]$, that is, all
causal effects can be readily estimated with an additive model. Our goal is to infer $\EE[Y | \Do(X_1=x)]$, based on $n=500$ and $n=10'000$  samples of the joint distribution of the $4$ nodes. The results are displayed in Figure~\ref{fig.non-add}.  

We consider the entire path-based Algorithm~\ref{alg1} (and
Algorithm~\ref{alg2} as well, not shown) assuming an additive structural
equation model as in (\ref{addSEM}). We impressively see that this approach
is exposed to model misspecification while \emph{S-mint} (in this case
simply fitting of an additive model, i.e., $b_{\mathrm{stop}} = 1$ with the
number of additional boosting iterations equaling zero) is not and leads to
reliable and correct results. We 
included two settings; $n=500$ to be consistent with the settings in the
numerical study from Section~\ref{sec.empirical.add} and $n=10000$ to
demonstrate that the failure of the path-based methods is not a small
sample size but an inconsistency phenomenon.

\subsection{Causal effects in the presence of backdoor paths} \label{subsec:bdpaths}
\begin{figure}[!htb]
\begin{center}
\includegraphics[width=0.73\textwidth]{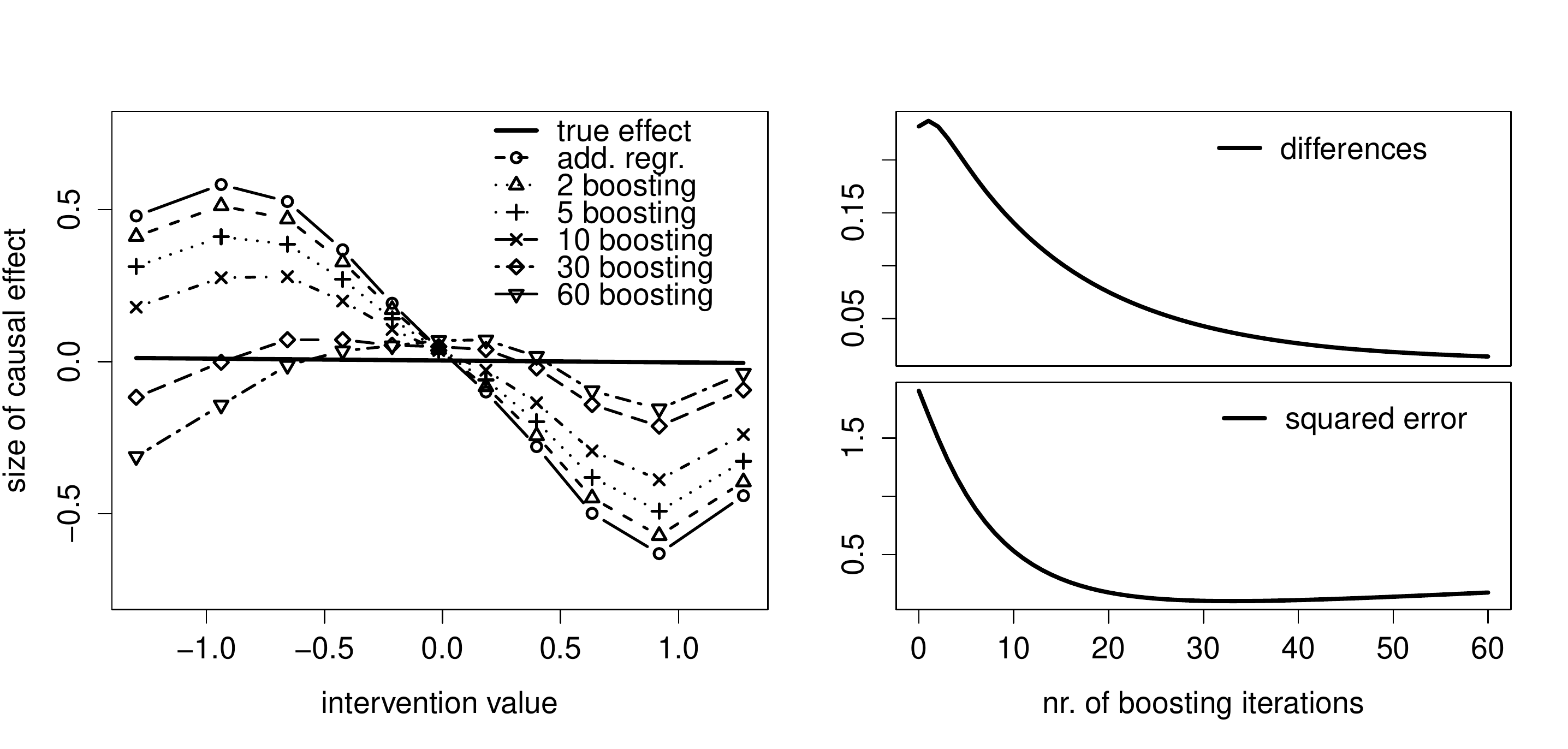} \\
\vspace{-0.5cm}
\includegraphics[width=0.73\textwidth]{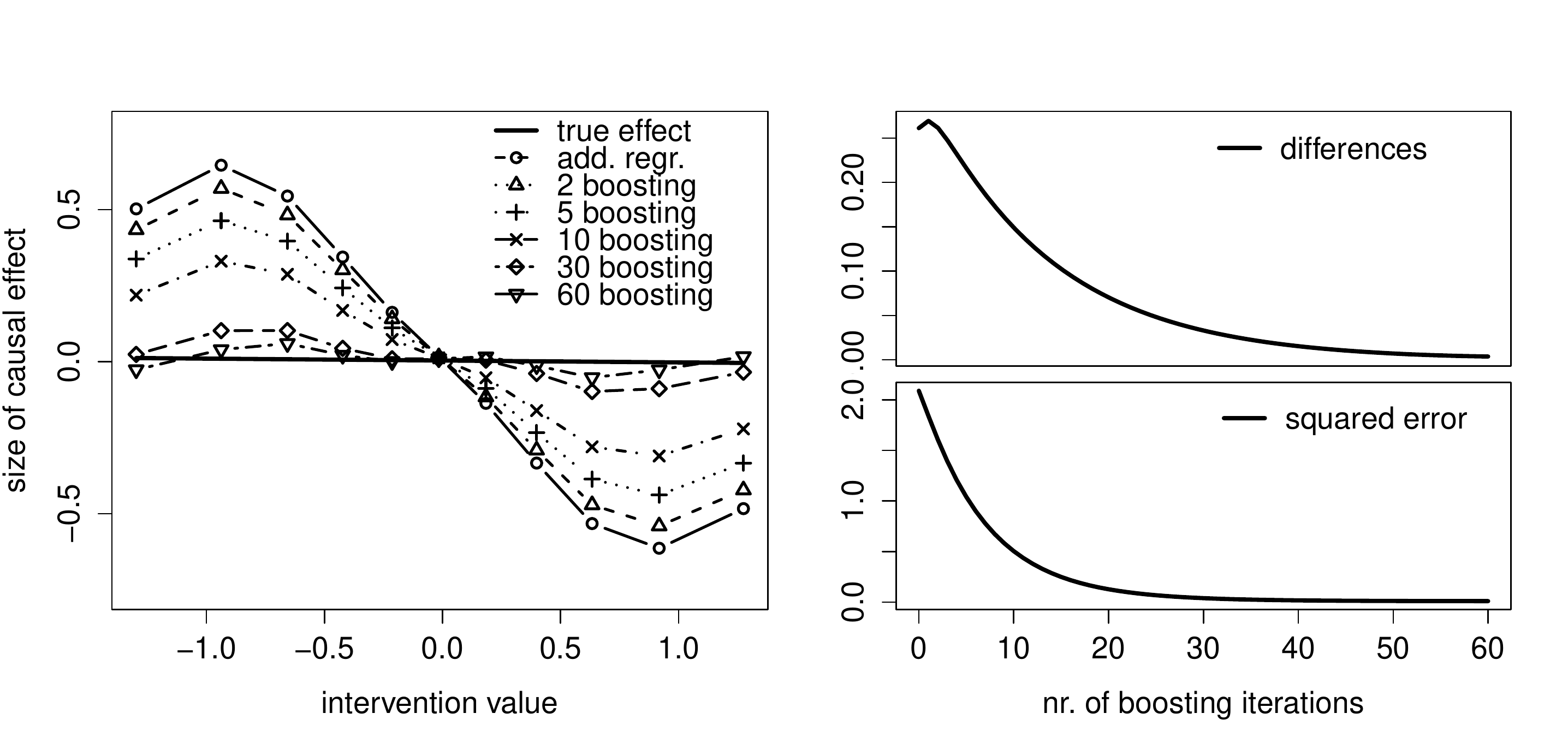}
\end{center}
\caption{Approximation of the causal effect $\EE[Y|\Do(X_3=x)]$ in model 
  (\ref{nonadd-SEM2}) with \emph{S-mint}
  regression for additive model fit (starting value)
  and various boosting iterations (left),
  absolute differences between consecutive boosting iterations as in
  (\ref{boostdiff}) (upper right) and integrated squared error for
  approximating the true 
  effect as a function of boosting iterations (lower right). The boosting
  iterations 
  in the \emph{S-mint} procedure account for interactions between the
  variables. The adjustment set is chosen as the 
  parental set of $X_3$, that is $S(j_X=3) = \{1,2\}$. The results are based
  on one representative
  sample of size $n=500$ (top) and $n=10000$ (bottom).}  
\label{fig:CEbdpaths}
\end{figure}

We now consider a slight (but crucial) modification of the above DAG that
has been proposed by Linbo Wang and Mathias Drton through private
communication. We consider the 4-node graph from Section~\ref{subsec:nobdpaths} with additional edges $X_1 \rightarrow Y$ and $X_2
\rightarrow Y$ and corresponding structural equation model 
\begin{align}\label{nonadd-SEM2}
	X_1 &\leftarrow \eps_1 \nonumber\\
	X_2 &\leftarrow \eps_2 \nonumber\\
	X_3 &\leftarrow X_1 + X_2 + \eps_3 \nonumber \\
	Y \  &\leftarrow X_1 \cdot X_2 \cdot X_3 + \eps_4
\end{align}
where $\eps_j \sim \mathcal{N}(0, \sigma_j^2)$ with $\sigma_1 = \sigma_2 = 0.7$ and $\sigma_3 = \sigma_4 = 0.2$. 
Note that this modification introduces two backdoor paths from $X_3$ to
$Y$. The goal is to estimate the causal effect $\EE[Y|\Do(X_3=x)]$ using
the \emph{S-mint} estimation procedure introduced in Algorithm~\ref{alg0}
with different numbers of boosting iterations. In Figure~\ref{fig:CEbdpaths} one clearly sees that the additive approximation (with
no additional boosting iterations) fails
to approximate the total causal effect.  
It is not able to capture the full
interaction term $X_1 \cdot X_2 \cdot X_3$. However, adding boosting
iterations significantly improves the approximation of the true causal
effect even for the small sample size $n=500$. 

\subsection{Causal effects in the presence of non-additive noise} 
\label{subsec:nonaddnoise}
\begin{figure}[!htb]
\begin{center}
\includegraphics[width=0.88\textwidth]{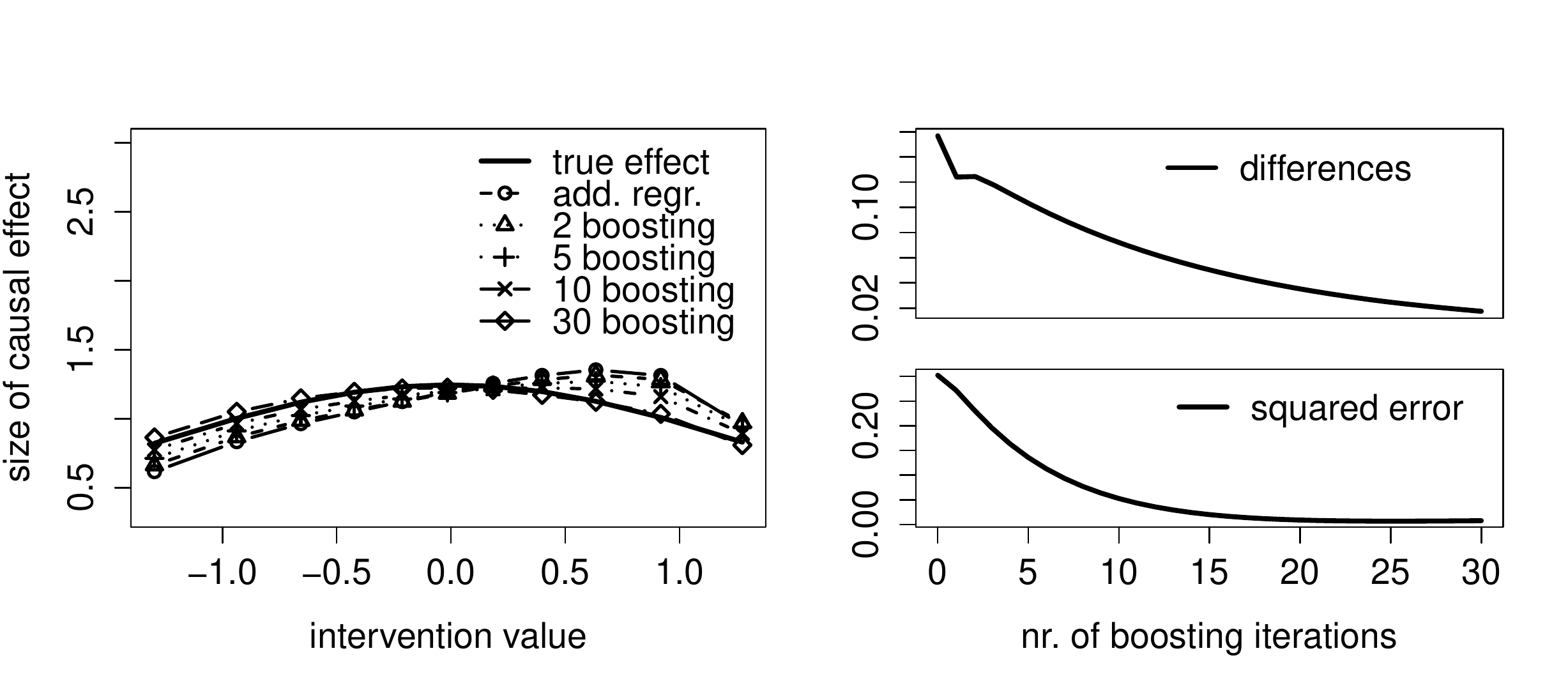} \\
\end{center}
\caption{Approximation of the causal effect $\EE[Y|\Do(X_3=x)]$ in model
  (\ref{SEMnonaddnoise}) exhibiting non-additive
  noise in the structural equation model, with \emph{S-mint} 
  regression for additive model fit (starting value)
  and various boosting iterations (left). Absolute differences between
  consecutive boosting iterations as in (\ref{boostdiff}) (upper right) and integrated squared error for 
  approximating the true effect as a function of boosting iterations
  (lower right). The adjustment set is chosen as the parental set of $X_3$,
  that is $S(j_X=3) = \{1,2\}$. The results are based on one representative
  sample of size $n=500$.}    
\label{fig:CEnonadd-noise}
\end{figure}

Theorem~\ref{th1} does not put any explicit restrictions on the noise
structure in 
the structural equation model. In particular, \emph{S-mint} also works well
in the case of non-additive noise. As an example, 
we consider the causal graph and SEM from Section~\ref{subsec:bdpaths} but replace
the structural equation corresponding to $Y$ in~\eqref{nonadd-SEM2} with
\begin{eqnarray}\label{SEMnonaddnoise}
Y \  \leftarrow \exp(X_1) \cdot \cos(X_2 \cdot X_3 + \eps_4).
\end{eqnarray}
The goal is again to estimate the causal effect $\EE[Y|\Do(X_3=x)]$ based on ${n=500}$ observed samples of the joint distribution.
Figure~\ref{fig:CEnonadd-noise} shows that \emph{S-mint} yields a close
approximation to the true causal effect. 

\subsection{Choice of the bandwidth}
\label{subsec:bandwidth}
\begin{figure}[!htb]
\begin{center}
\includegraphics[width=0.88\textwidth]{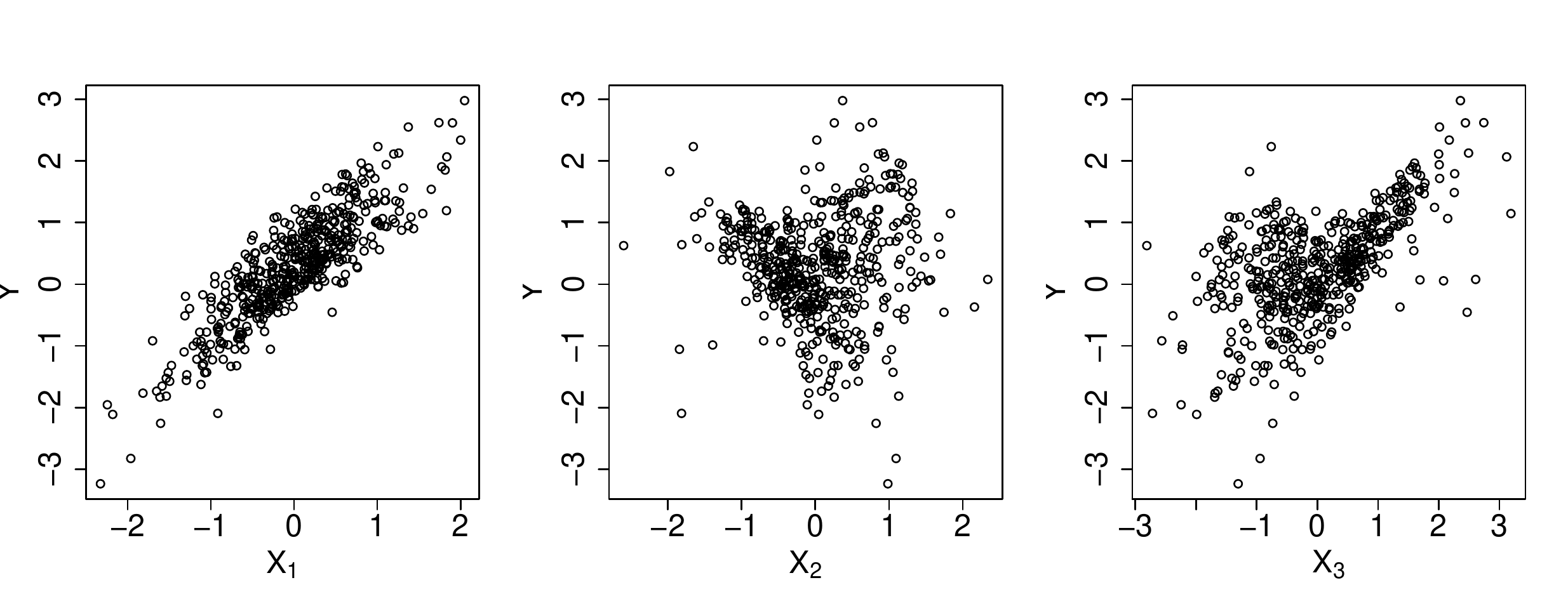} \\
\end{center}
\caption{Scatterplots of the data from model (\ref{SEM-bandwidthill}) of
  $Y$ versus $X_1,X_2$ and $X_3$. They reveal a difference in wigglyness.}  
\label{fig:scatter}
\end{figure}

\begin{figure}[!htb]
\begin{center}
\includegraphics[width=0.88\textwidth]{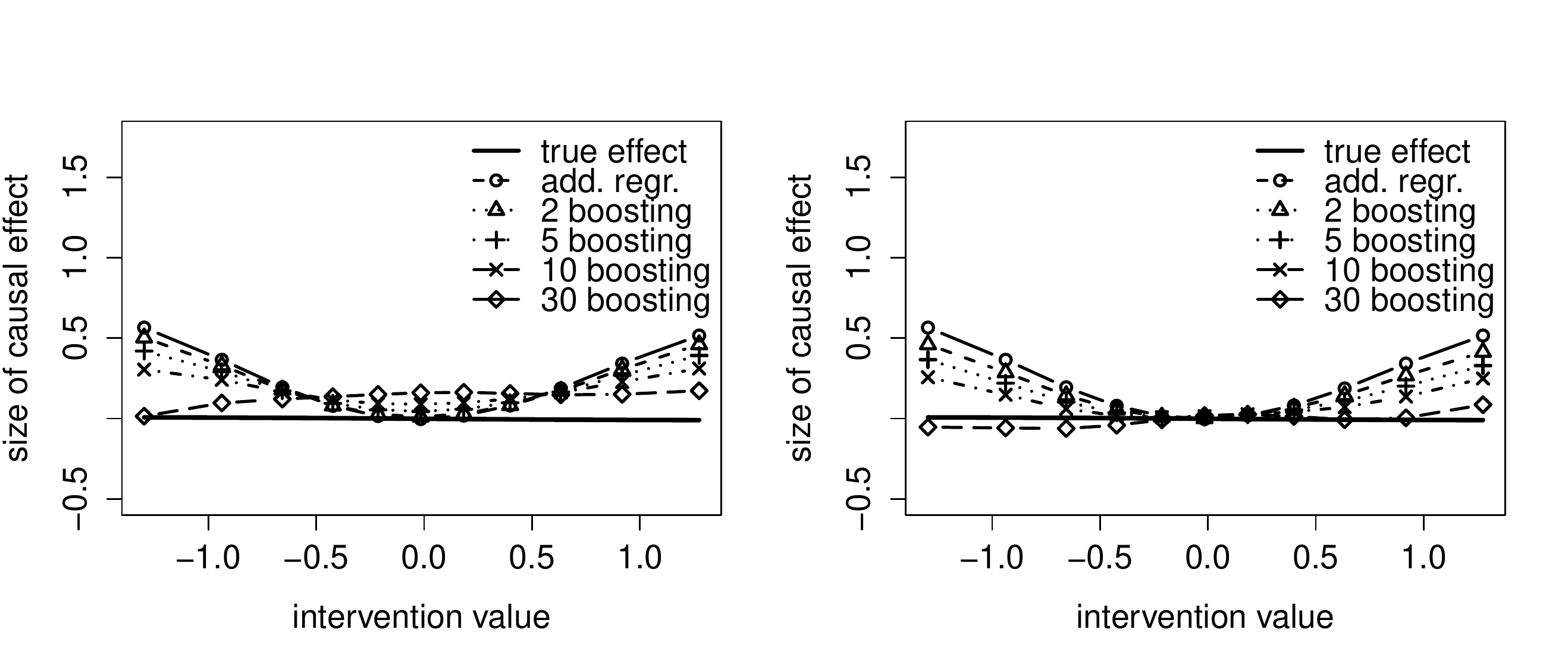} \\
\end{center}
\caption{Approximation of the causal effect $\EE[Y|\Do(X_3=x)]$ in model
  (\ref{SEM-bandwidthill}). The
  adjustment set is chosen as the parental set of $X_3$, that is $S(j_X=3) =
  \{1,2\}$ and with corresponding fixed bandwidths
  $h_2^{(1)}=h_2^{(2)}=0.4$ (left) and varying bandwidths $h_2^{(1)}=0.8$
  and $h_2^{(2)}=0.4$ (right). The results are based on one representative
  sample of size $n=500$.}  
\label{fig:CEbdpaths_sin}
\end{figure}

Theorem~\ref{th1} provides an asymptotic result but does not specify how to
choose the bandwidths $h_1$ and $h_2$ in the finite sample case. In
particular, the same fixed choice of $h_2$ for all variables in the
adjustment set $S$ can be suboptimal in some situations. As an example let us
consider the graph and structural equations from
Section~\ref{subsec:bdpaths} where we replace one equation
in~\eqref{nonadd-SEM2} by 
\begin{eqnarray}\label{SEM-bandwidthill}
	Y \  \leftarrow X_1 + \sin(X_2 \cdot X_3 ) + \eps_4.
\end{eqnarray}
The goal is to approximate the causal effect $\EE[Y|\Do(X_3=x)]$ based on
$n=500$ samples of the joint distribution. Inspecting the scatterplots of
$Y$ versus $X_1, X_2$ and $X_3$ (see Figure~\ref{fig:scatter}) suggests
that the bandwidth $h_{2}^{(1)}$ corresponding to $X_1$ should be larger
than the bandwidth $h_2^{(2)}$ corresponding to $X_2$. 
Figure~\ref{fig:CEbdpaths_sin} depicts the corresponding approximated
causal effects using the \emph{S-mint} method for a fixed bandwidth $h_2 =
(h_2^{(1)}, h_2^{(2)}) = (0.4,0.4)$ and for a variable bandwidth $h_2 =
(h_2^{(1)}, h_2^{(2)}) = (0.8,0.4)$ respectively. Clearly, the
approximation with the variable bandwidth outperforms the approximation with
the fixed bandwidth. Adaptive bandwidths choice methods as proposed by
\citet{polzehl2000adaptive} might be suitable, at the price of a more
complicated and hence more variable estimation scheme. 

\section{Empirical results: additive structural equation models}\label{sec.empirical.add}

The goal of the numerical experiments in this section is to quantify the
estimation accuracy of the total causal effect $\EE[Y|\Do(X=x)]$
for two variables $X,Y \in 
\{X_1,...,X_p\}$ such that $Y$ is a descendant of $X$ (if $Y$ is an
ancestor of $X$, then the interventional expectation corresponds to the
observational expectation $\EE[Y]$). We consider in this section only
additive structural equation models as in~\eqref{addSEM}. This allows for a comparison of the \emph{S-mint} method and the path-based methods.

For \emph{S-mint} regression, we use the implementation described in Section~\ref{subsec.implem}. The kernel functions $K$ and $L$ in the \emph{S-mint} procedure are chosen to be a Gaussian kernel with bandwidth $h_1$ and a product of Gaussian kernels with bandwidth $h_2$ respectively. 
For simplicity, in the style of \cite{fan1998direct}, we choose $h_1$ and $h_2$ as $0.5$ times the empirical standard deviation of the respective covariables in all of our simulations in this section. 
We use the following two criteria for $b_{\text{stop}}$, that is, as an automated stopping criterion for the boosting iterations: 
\begin{enumerate}
\item Stop if an iteration changes the approximation by less than $1\%$. That is, the integrated difference (\ref{boostdiff}) to the previous approximation is less than~$0.01$.
\item Stop if the integrated difference between two consecutive approximations is less than $5\%$ of the initial integrated difference.
\end{enumerate}

When using the path-based methods from Section~\ref{sec.CAMpath}, we
estimate the functions $f^0_j$ by additive functions using the
\texttt{R}-package \texttt{mgcv} with default values (and thus using the
knowledge of the form of the nonlinear functions in the SEM).  

We test the performance of four different methods: \emph{S-mint} with
parental sets (Algorithm~\ref{alg0}) with the stopping of boosting
iterations as described above, additive regression with parental
sets (first step of \emph{S-mint}, without additional boosting iterations),
entire path-based method from root nodes (Algorithm~\ref{alg1}) and
partially path-based method with short-cuts  
(Algorithm~\ref{alg2}).
The reference effect $\EE[Y|\Do(X=x)]$ is computed using
Algorithm~\ref{alg1} with known (true) functions $f_{j,k}^0$ and error
variances $(\sigma^0_j)^2$ based on $5 n$ samples.  

Since in a nonlinear structural equation model (in contrast to a
linear structural equation model) $\EE[Y|\Do(X=x)]$ is a nonlinear function
of the intervention value $x$, we compute the
interventional expectation for several values $x$: typically, for the nine
deciles $d_1(X),...,d_9(X)$ of $X$.
To compare the estimation accuracy of the three methods on DAG $D$, we
compute a relative squared error $e(D)$ over all considered pairs $(X,Y)$
(for details see below), denoted by $\mathcal{L}$, and over all
intervention values 
$d_1(X),...,d_9(X)$ as    
\begin{equation} \label{eq:approxError}
	e(D) = \frac{\sum\limits_{(X,Y) \in \mathcal{L}}
          \,\,\sum\limits_{i=1}^9 \left(\hat{\EE}[Y|\Do(X=d_i(X))] -
            \EE^0[Y|\Do(X=d_i(X))] \right)^2}{\sum\limits_{(X,Y) \in
            \mathcal{L}} \,\, \sum\limits_{i=1}^9
          \left(\EE^0[Y|\Do(X=d_i(X))] \right)^2 }. 
\end{equation} 

Typically, we repeat every experiment on $N=50$ or $N=100$ random DAGs
(described in Section~\ref{ssec:datasim}) and
record the relative error $e(D)$ of all methods for each repetition.

\subsection{Data simulation} \label{ssec:datasim}
To simulate data we first fix a causal order $\pi^0$ of the variables,
that is \linebreak $X_{\pi^0(1)} \prec X_{\pi^0(2)} \prec \cdots \prec X_{\pi^0(p)}$ and
include each of the $\tbinom{p}{2}$ possible directed edges, independently
of each other, with
probability $p_c$. In the sparse setting we typically choose
$p_c=\frac{2}{p-1}$ which yields an expected number of $p$ edges in the
resulting DAG. Based on the causal structure of the graph we then build the
structural equation model. We simulate from the
additive structural equation model \eqref{addSEM}, where every edge $k
\rightarrow j$ in the DAG is associated with a nonlinear function
$f^0_{j,k}$ in the structural equation model. We
use two function types:
\begin{enumerate}
\item edge functions $f_{j,k}^0$ drawn from a Gaussian process with a Gaussian kernel with bandwidth one
\item sigmoid-type edge functions of the form $f_{j,k}^{0}(x)=a \cdot \frac{b \cdot
  (x+c)}{1+|b \cdot (x+c)|}$ with ${a \sim \text{Exp}(4)+1}$, $b \sim
\text{Unif}([-2,-0.5]\cup[0.5,2])$ and $c \sim \text{Unif}([-2,2])$.
\end{enumerate}
All variables with empty parental set (root nodes in the DAG) follow a
Gaussian distribution with mean zero and standard deviation which is
uniformly distributed in the interval $[1,\sqrt{2}]$. To all remaining
variables we add Gaussian noise with standard deviation uniformly
distributed in $[1/5, \sqrt{2}/5]$. Note that both simulation settings
correspond to the ones used by \citet{pbjoja13}. 

\subsection{Estimation of causal effects with known graphs} \label{ssec:CEknownDAG}
In this section we compare the different methods in terms of estimation
accuracy and CPU time consumption for known underlying DAGs $D^0$. To that
end we generate random DAGs with $p=10$ variables and simulate $n=500$
samples of the joint distribution applying the simulation procedure
introduced in Section~\ref{ssec:datasim}. We then select all
index pairs $(k,j)$ such that there exists a directed path from $X_k$ to
$X_j$ and estimate the causal effect $\EE[X_j|\Do(X_k)]$ for all $k,j$
on the nine deciles of $X_k$. 

\begin{figure}[h]
\begin{center}
\includegraphics[width=0.48\textwidth]{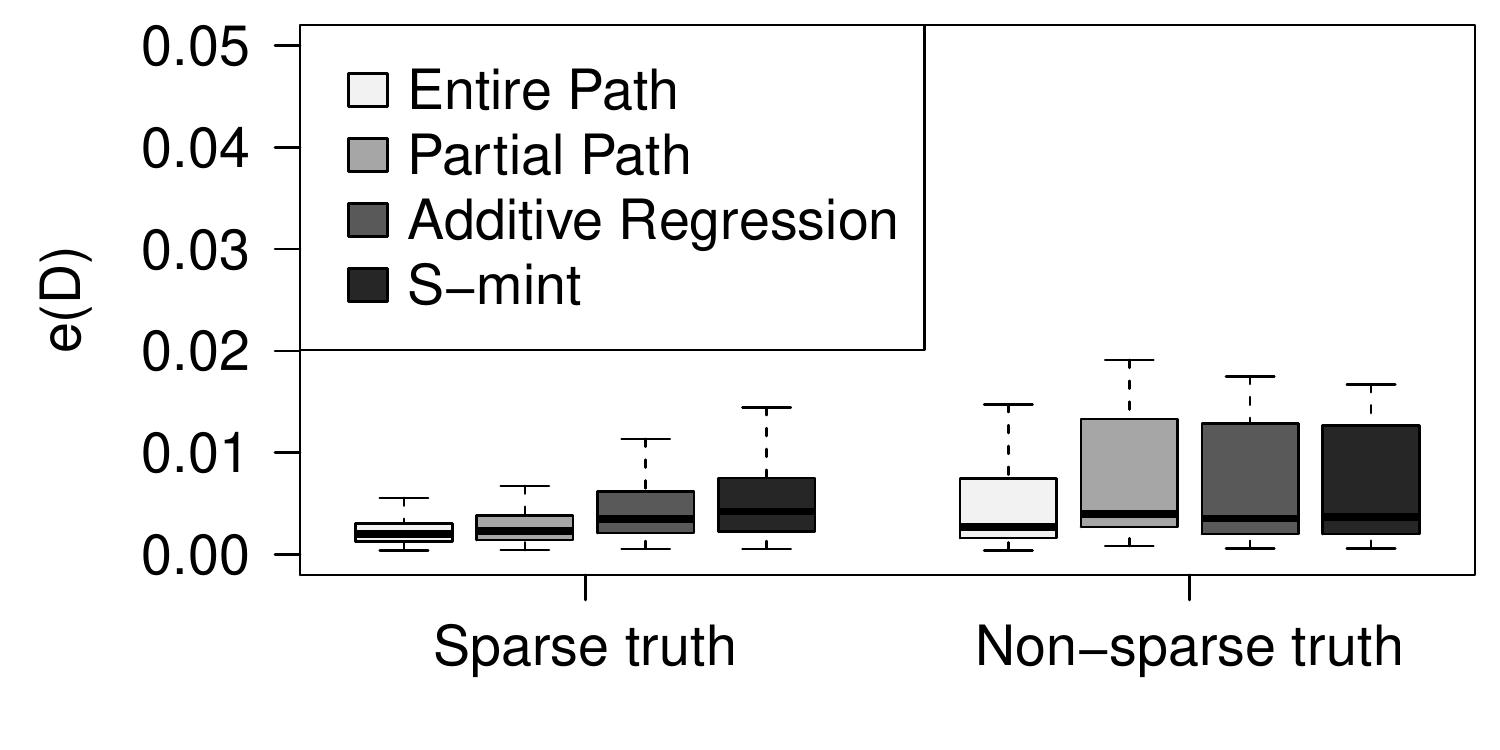}
\hfill 
\includegraphics[width=0.48\textwidth]{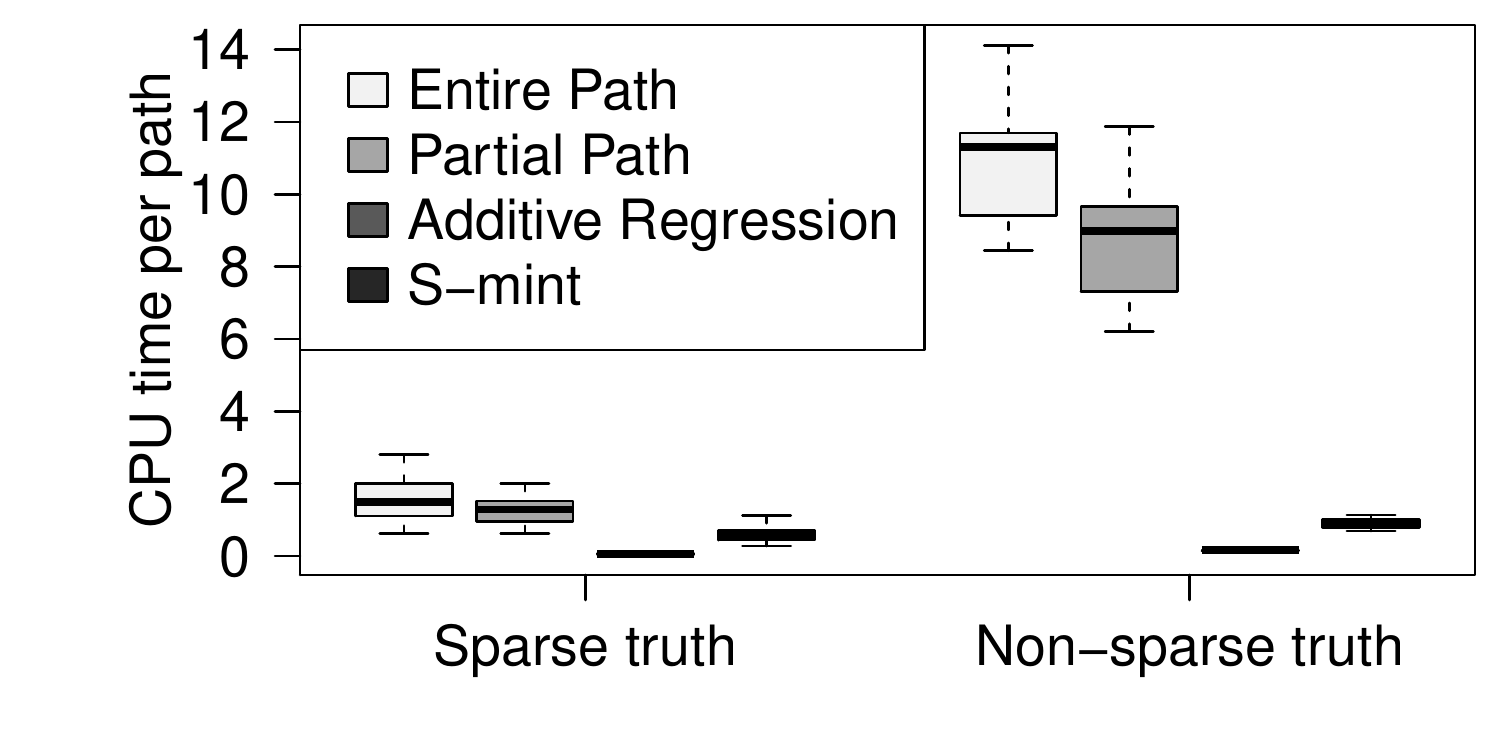} 
\end{center}
\caption{Comparison of the performance of the methods in terms of relative
  squared error as in (\ref{eq:approxError}) (left) and CPU time
  consumption (right) for the case where 
  the true DAGs $D^0$ are known and the edge functions belong to the
  sigmoid-type setting. The adjustment set is $S=\pa(X_k)$ for additive
  regression and \emph{S-mint}. Number of variables $p=10$ and sample size
  $n=500$.} 
\label{fig:CEknownDAG}
\end{figure}

\begin{figure}[h]
\begin{center}
\includegraphics[width=0.48\textwidth]{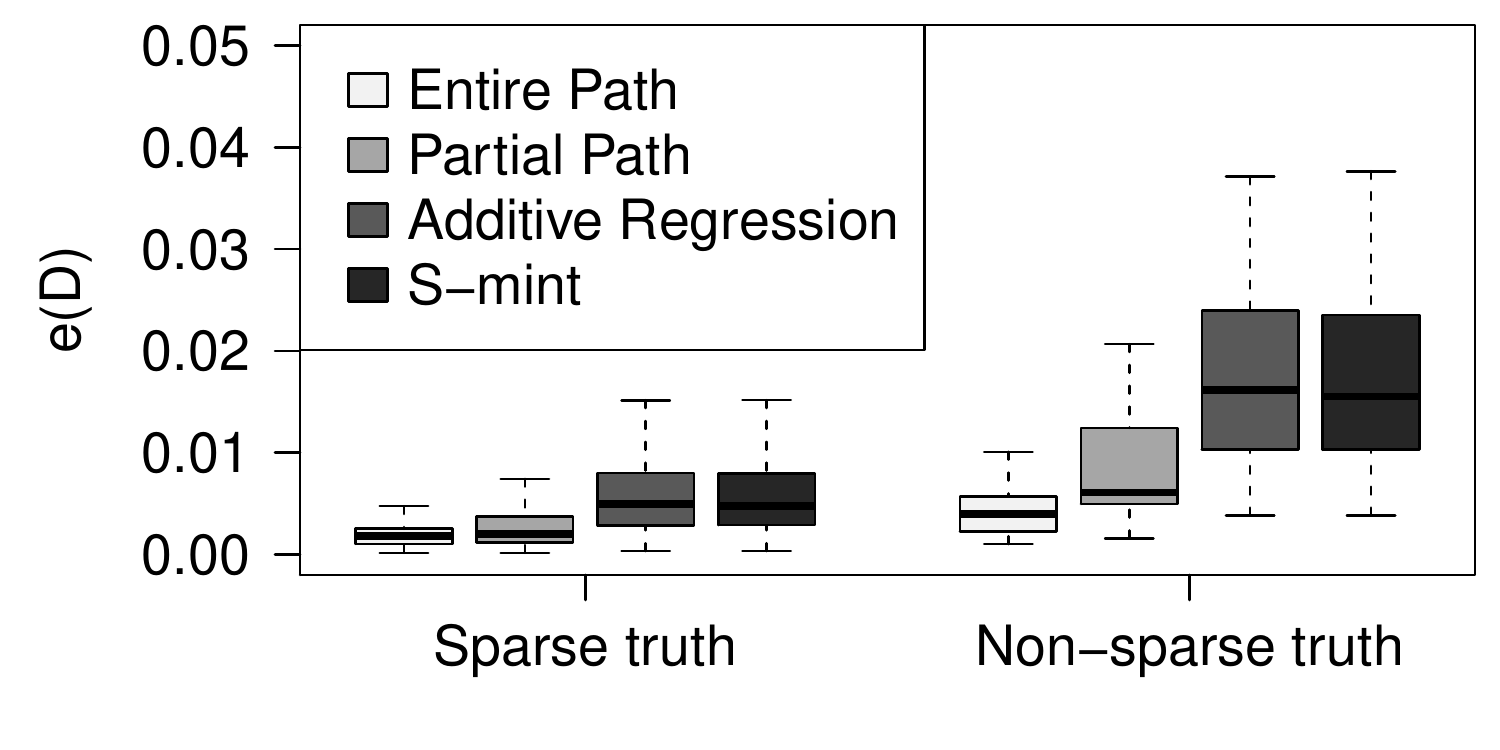}  
\hfill 
\includegraphics[width=0.48\textwidth]{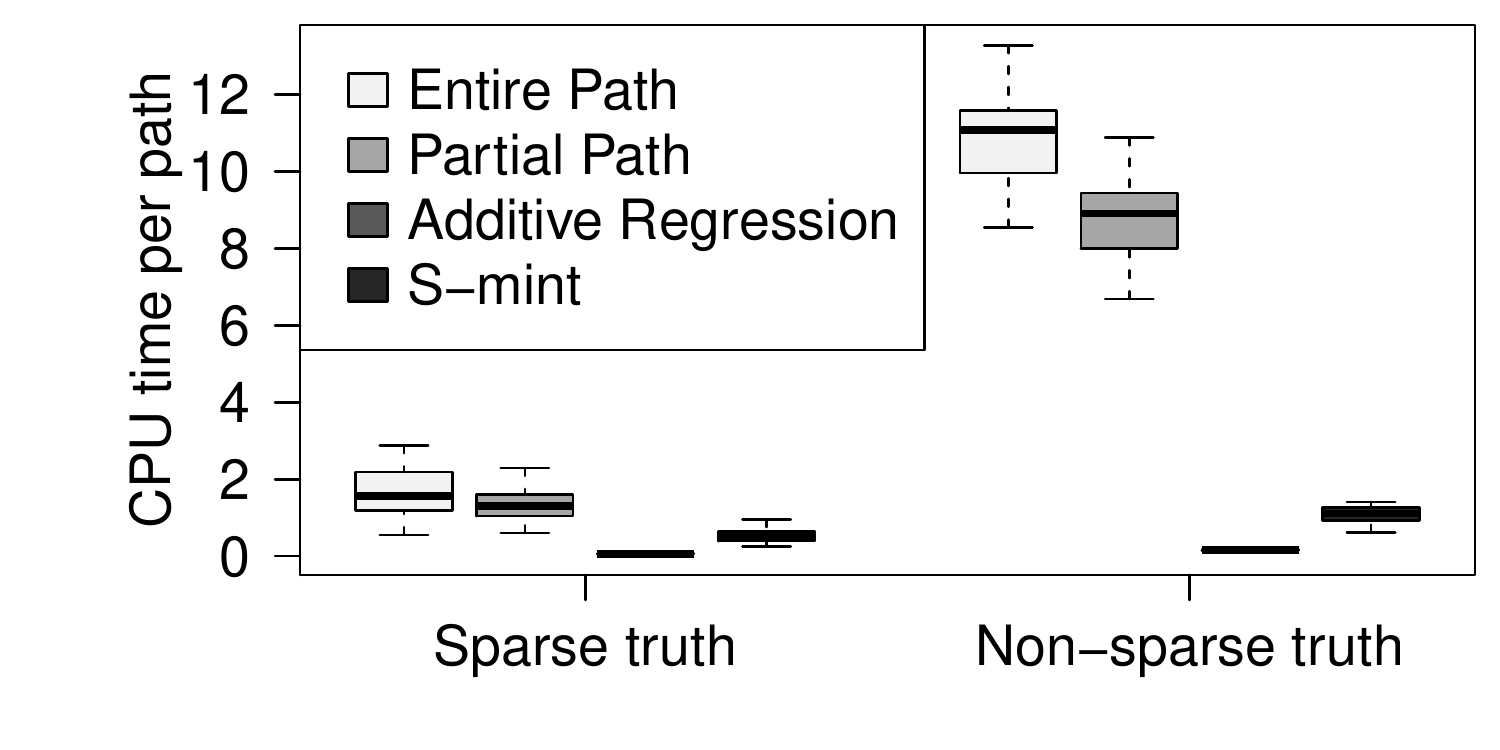} 
\end{center}
\caption{Comparison of the performance of the methods in terms of relative
  squared error as in (\ref{eq:approxError}) (left) and CPU time consumption
  (right) for the case where 
  the true DAGs $D^0$ are known and the edge functions are drawn from a
  Gaussian process with bandwidth one. The adjustment set is $S=\pa(X_k)$
  for additive regression and \emph{S-mint}. Number of variables $p=10$ and
  sample size $n=500$.} 
\label{fig:CEknownDAG.GAM}
\end{figure}

The experiment is done for two different levels of sparsity, a sparse graph with an
expected number of $p$ edges and a non-sparse graph with an expected number
of $4p$ edges. We record the relative squared error \eqref{eq:approxError}
and the CPU time consumption, both 
averaged over all index pairs, for $N=100$ ($N=20$ in the dense setting,
respectively) different DAGs $D^0$. The results are displayed in
Figure~\ref{fig:CEknownDAG} for the sigmoid-type edge functions and in
Figure~\ref{fig:CEknownDAG.GAM} for the Gaussian process-type edge
functions.  

The method based on the entire paths (Algorithm~\ref{alg1}) yields the
smallest errors followed by the path-based methods with short-cuts
(Algorithm~\ref{alg2}). The \emph{S-mint} and additive regression exhibit a
slightly worse performance. This finding can be explained by the fact that
the path-based methods benefit from the full (and correct) structural
information of the DAG whereas the \emph{S-mint} and additive regression
methods only use local information (cf. Section~\ref{subsec.local}). For
the monotone sigmoid-type function class, additive regression 
provides a
very good approximation to the true causal effect even in dense
settings. 
For both settings we observe that the boosting iterations in \emph{S-mint} do not improve the additive approximation substantially.

In terms of CPU time consumption, \emph{S-mint} and additive regression
outperform the path-based methods. Additive regression is particularly
fast as it only requires the fit 
of one nonparametric additive regression of $X_j$ versus $X_k$ and
$X_{\pa(k)}$ whereas the path-based methods each require one nonparametric
additive model fit for every node on all the traversed paths. As the set of
paths in the partially path-based method is a subset of the one in the
entire path-based method (cf. Section~\ref{subsec.partial} and Figure~\ref{fig:illustAlgorithms2}), the
partially path-based method needs less model fits which explains the
reduction of time consumption. In particular, both \emph{S-mint} and
additive regression are computationally feasible for computing $\EE[X_j
|\Do(X_k)]$ for all pairs $(k,j)$, even when $p$ is large and in the
thousands assuming that the cardinality of the corresponding adjustment
sets is reasonably small.  

\subsection{Estimation of causal effects on perturbed graphs} \label{ssec:CEmodifDAG}

\begin{figure}[h]
\begin{center}
\includegraphics[width=0.76\textwidth]{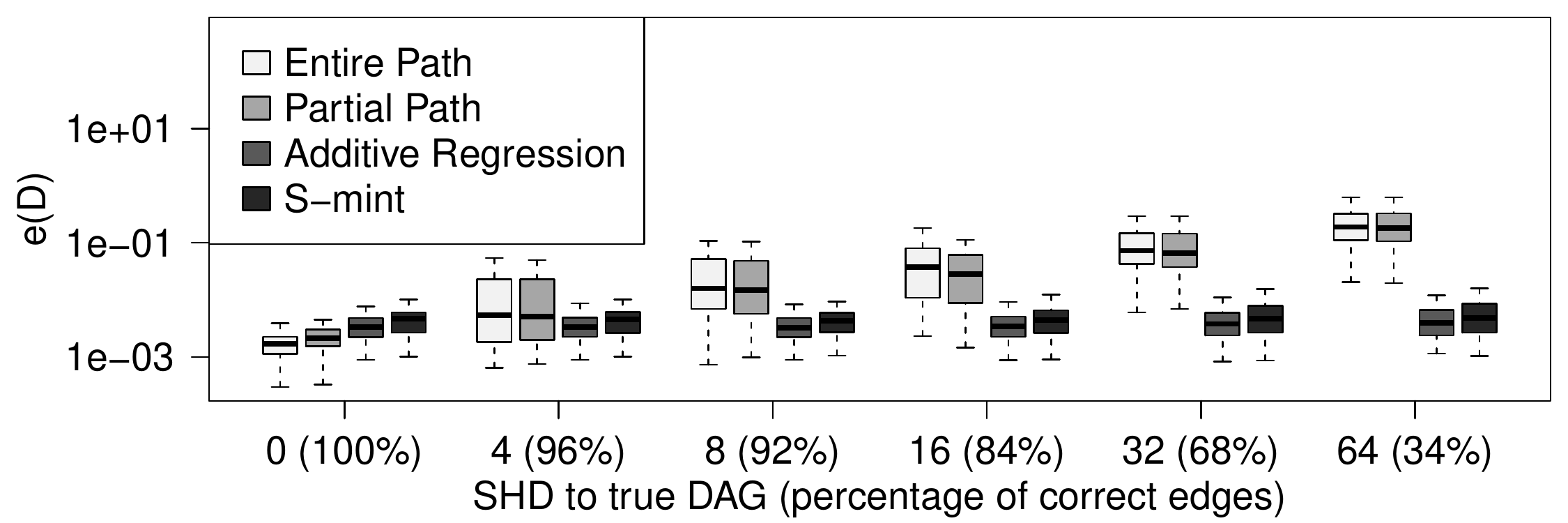}  \\ 
\includegraphics[width=0.76\textwidth]{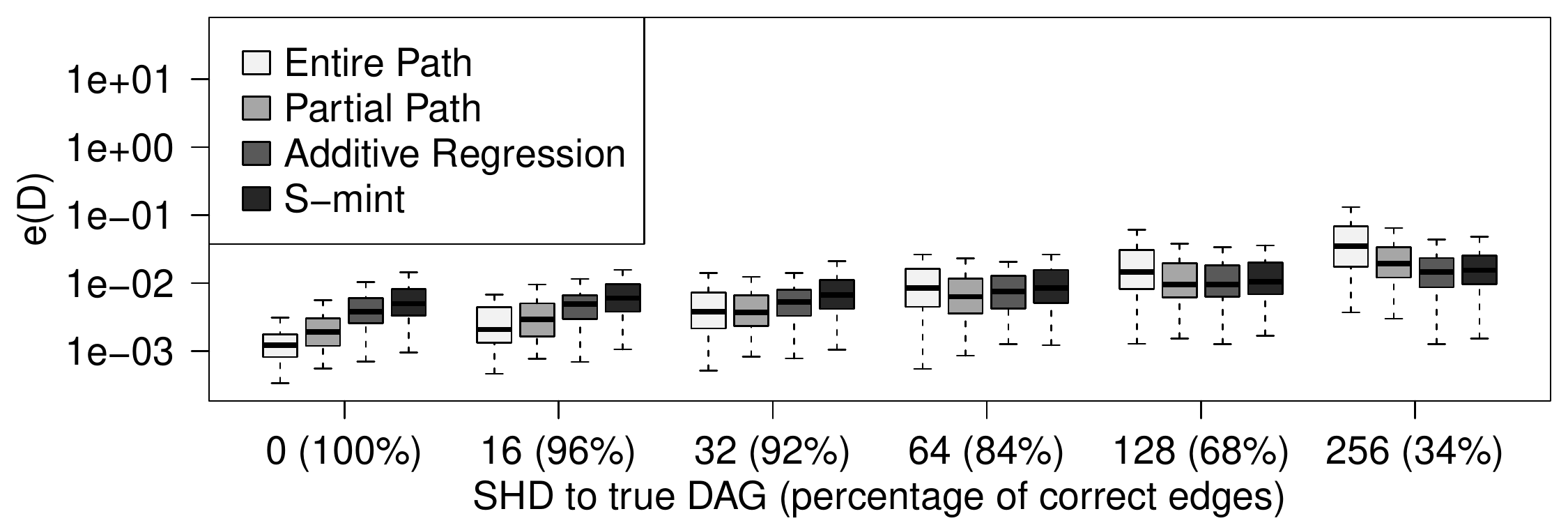} 
\end{center}
\caption{The plots
  compare the relative squared error performance of the three 
  methods on a set of modified DAGs $\{\tilde{D}_{h_r}\}_{r \in \mathcal{K}}$ with given
  structural Hamming distances $\{h_r\}_{r \in \mathcal{K}}$ to the true
  DAG $D^0$ (or equivalently, with a given percentage of correct edges) for the sigmoid-type additive structural equation model. The
  top and bottom panels show the relative squared error
  error $e(D)$ (\ref{eq:approxError}) in a sparse and dense
  setting, respectively. The larger the structural Hamming distance $h_r$
  between the modified DAG $\tilde{D}_{h_r}$ and the true DAG $D^0$, the
  better is the performance of \emph{S-mint} with parental sets
  in comparison with the two path-based methods. Number of variables $p=50$
  and sample size $n=500$.} 
\label{fig:CEmodifDAG}
\end{figure}

\begin{figure}[h]
\begin{center}
\includegraphics[width=0.76\textwidth]{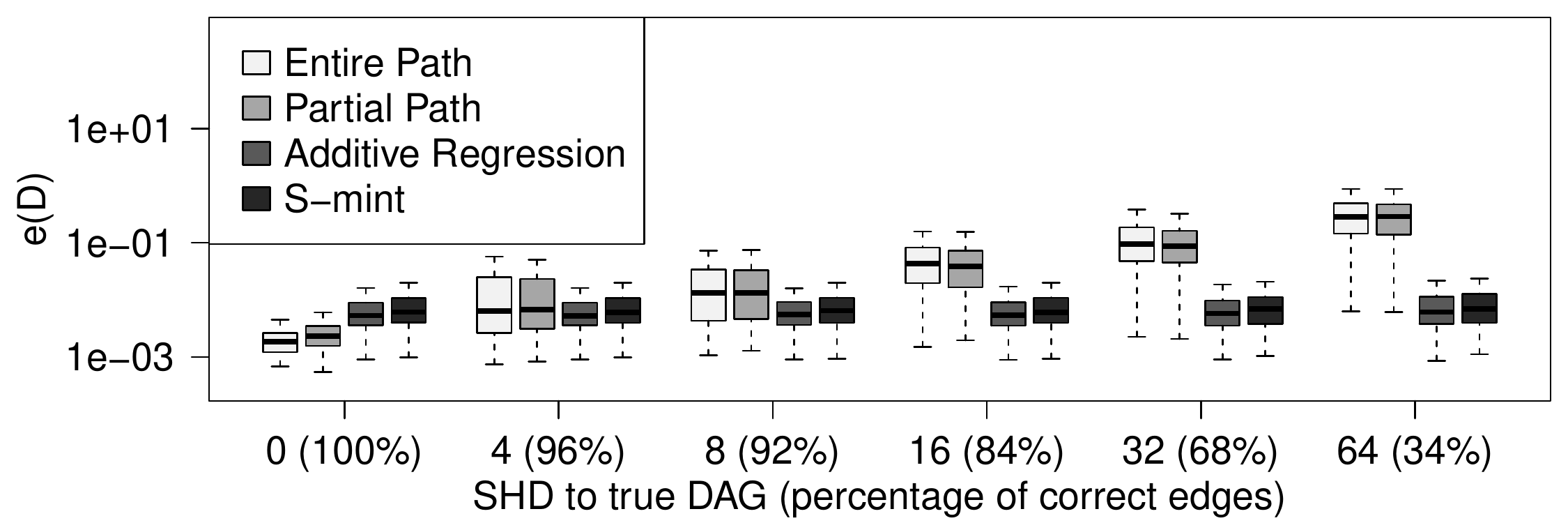}  \\ 
\includegraphics[width=0.76\textwidth]{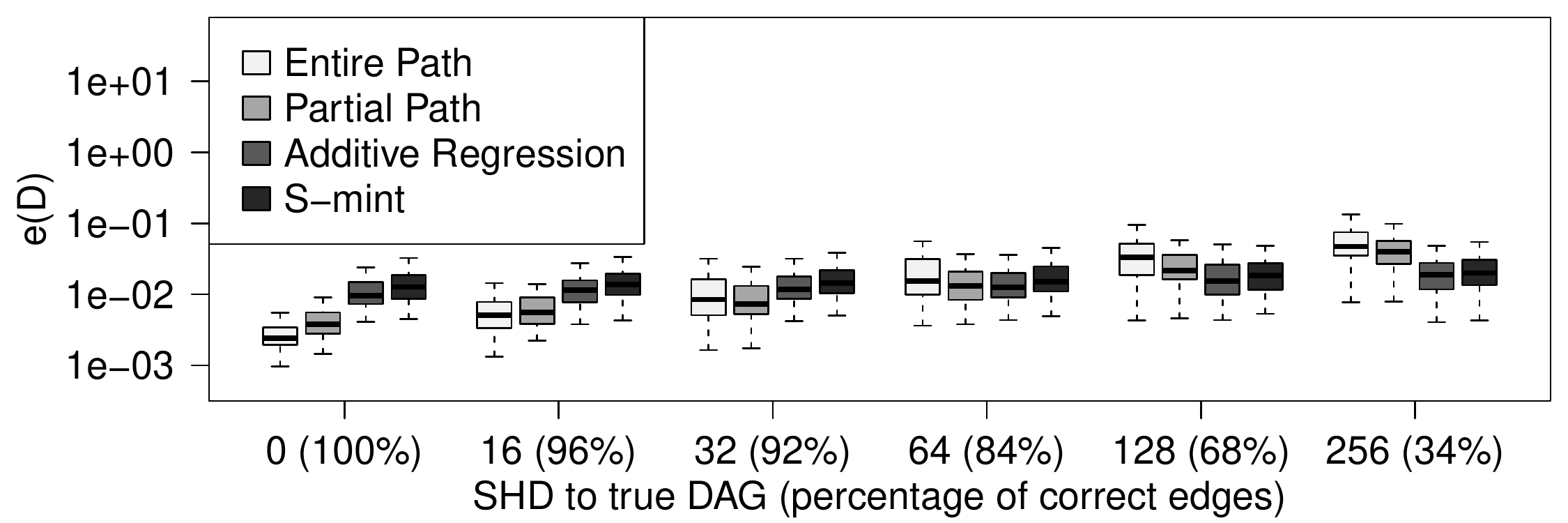}  
\end{center}
\caption{The plots compare the relative squared error performance of the three
  methods on a set of 
  modified DAGs $\{\tilde{D}_{h_r}\}_{r \in \mathcal{K}}$ with given
  structural Hamming distances $\{h_r\}_{r \in \mathcal{K}}$ to the true
  DAG $D^0$ (or equivalently, with a given percentage of correct edges) for the Gaussian process-type additive structural equation models. The
  top and bottom panels show the relative squared error
  $e(D)$ (\ref{eq:approxError}) in a sparse and dense 
  setting, respectively. The larger the structural Hamming distance $h_r$
  between the modified DAG $\tilde{D}_{h_r}$ and the true DAG $D^0$, the
  better is the performance of \emph{S-mint} with parental sets
  in comparison with the two path-based methods. Number of variables $p=50$
  and sample size $n=500$.} 
\label{fig:CEmodifDAG.GAM}
\end{figure}

In the previous section we demonstrated that the two path-based methods
exhibit a better performance than \emph{S-mint} and the additive regression
approximation if causal 
effects are estimated based on the underlying true DAG $D^0$. We will now
focus on the more realistic situation in which we are only provided with a
partially correct DAG $\tilde{D}$. We model this by constructing a set of
modified DAGs $\{\tilde{D}_{h_r}\}_{r \in \mathcal{K}}$ with pre-specified
(fixed) structural Hamming distances $\{h_r\}_{r \in \mathcal{K}}$ to the
true DAG $D^0$, where ${\mathcal{K} = \{1,2,\ldots ,6\}}$ and the
corresponding $\{h_r\}_{r \in \mathcal{K}}$ are described in
Figures~\ref{fig:CEmodifDAG}~and~\ref{fig:CEmodifDAG.GAM}. To do so, we use
the following rule: starting from $D^0$ with $p=50$ nodes, for each $r \in
\mathcal{K}$, we randomly remove and add $\frac{h_r}{2}$ edges each to
obtain $\tilde{D}_{h_r}$. The structural Hamming distance between $D^0$ and
the perturbed graph $\tilde{D}_{h_r}$ is then equal to $h_r$, and a percentage of $1 -\frac{h_r}{2 |E|}$ edges in $\tilde{D}_{h_r}$ are still
correct, where $|E|$ denotes the expected number of edges in the DAG $D^0$. Note
that this modification may change the order of the variables (especially
for large values of $h_r$). 

We randomly
choose $20 = |\mathcal{L}|$ index pairs $(k,j)$ such that there exists a
directed path from 
$X_k$ to $X_j$ in $D^0$, but now consider the problem of estimating the total
causal effect $\EE[X_j|\Do(X_k)]$ based on the
perturbed graph $\tilde{D}_{h_r}$ for the adjustment sets or the paths,
respectively (and based on sample size $n=500$ as in 
Section~\ref{ssec:CEknownDAG}). For every $r \in \mathcal{K}$, this is
repeated $N=100$ times and in each repetition, we record the relative squared error $e(D)$ in \eqref{eq:approxError}. As before, we
distinguish between a sparse graph with an expected number of $50$ edges
and a non-sparse graph with an expected number of $200$ edges and we use
both simulation settings described in Section~\ref{ssec:datasim} for
generating the edge functions $f^0$. The results are shown in
Figures~\ref{fig:CEmodifDAG}~and~\ref{fig:CEmodifDAG.GAM}.   

For both, the sparse and non-sparse settings, one observes that the larger
the structural Hamming distance (or equivalently, the smaller the
percentage of correctly specified edges in $D^0$), the better is the
performance of \emph{S-mint} and additive regression in comparison with the path-based methods. That is, both methods are substantially more robust with
respect to possible misspecifications of edges in the graph.   
This may be explained by the different degrees of localness (cf. Section~\ref{subsec.local}) of the respective methods. For the two local methods 
we can hope to have approximately correct information in the 
parental set of $X_k$ even if the modified DAG is far away from the true
DAG $D^0$ in terms of the structural Hamming distance. For the path-based
methods however, randomly removing edges may break one or several of the
traversed paths which results in causal information being partially or
fully lost. This effect is most evident in the two sparse settings. A similar behavior is also observed in Figure~\ref{fig:CEestimDAG}. 

Note that except for the true DAG $D^0$, the performance of the partially path-based method is at least as good as for the entire path-based method. The shortcut introduced in Algorithm~\ref{alg2} does not only 
yield computational savings but also improves (relative to the full path-based Algorithm~\ref{alg1}) statistical estimation accuracy of causal effects in incorrect 
DAGs. Again, a possible explanation for this observation is that the
partially path-based method acts more locally and thus is less affected by
edge perturbations.

\subsection{Estimation of causal effects in estimated graphs} \label{ssec:CEestimDAG}

We now turn our attention to the case where the goal is to compute causal
effects on a DAG $\hat{D}$ that has been estimated by a structure learning
algorithm (while still relying on a correct model specification). In
conjunction with \emph{S-mint} regression, this is then the 
\emph{est S-mint} method described in Section~\ref{subsec.twostage}. 

We generate $N=50$ random DAGs with $p=20$ nodes for different numbers $n$ of
observational data, which are simulated according to the procedure in
Section~\ref{ssec:datasim}.  

Using the knowledge that the structural equation model is additive as in
(\ref{addSEM}), we 
apply the recently proposed CAM method \citep{pbjoja13} for estimation
of the true underlying DAG $D^0$ (which is identifiable from the
observational distribution), outlined at the end of Section
\ref{subsec.DAGinfer}. The implementation is according to the \texttt{R}-package \texttt{CAM}. Regarding the algorithmic details, we use the
following in the three steps: 
\begin{enumerate}
	\item Preliminary neighborhood selection to restrict the number of
          potential parents per node: set to a maximum of $10$ by default;
	\item Estimation of the correct order by greedy search: we use $6$
          basis functions per parent to fit the additive
          model; 
	\item Optional: Pruning of the DAG by feature selection to keep
          only the significant edges, where we use the default level
          $\alpha=0.001$. 
\end{enumerate}

\begin{figure}[h]
\begin{center}
\includegraphics[width=0.48\textwidth]{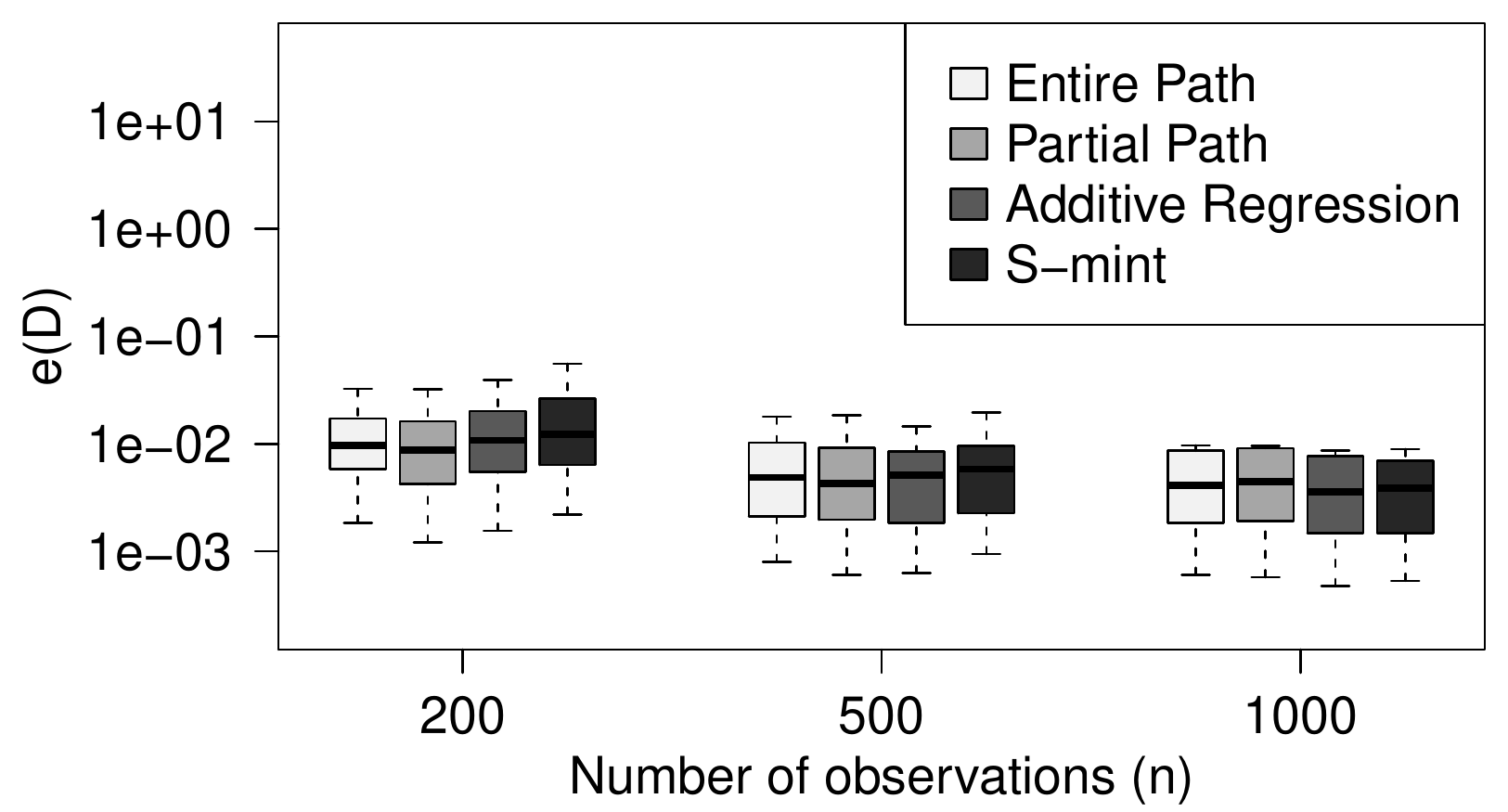}
\hfill
\includegraphics[width=0.48\textwidth]{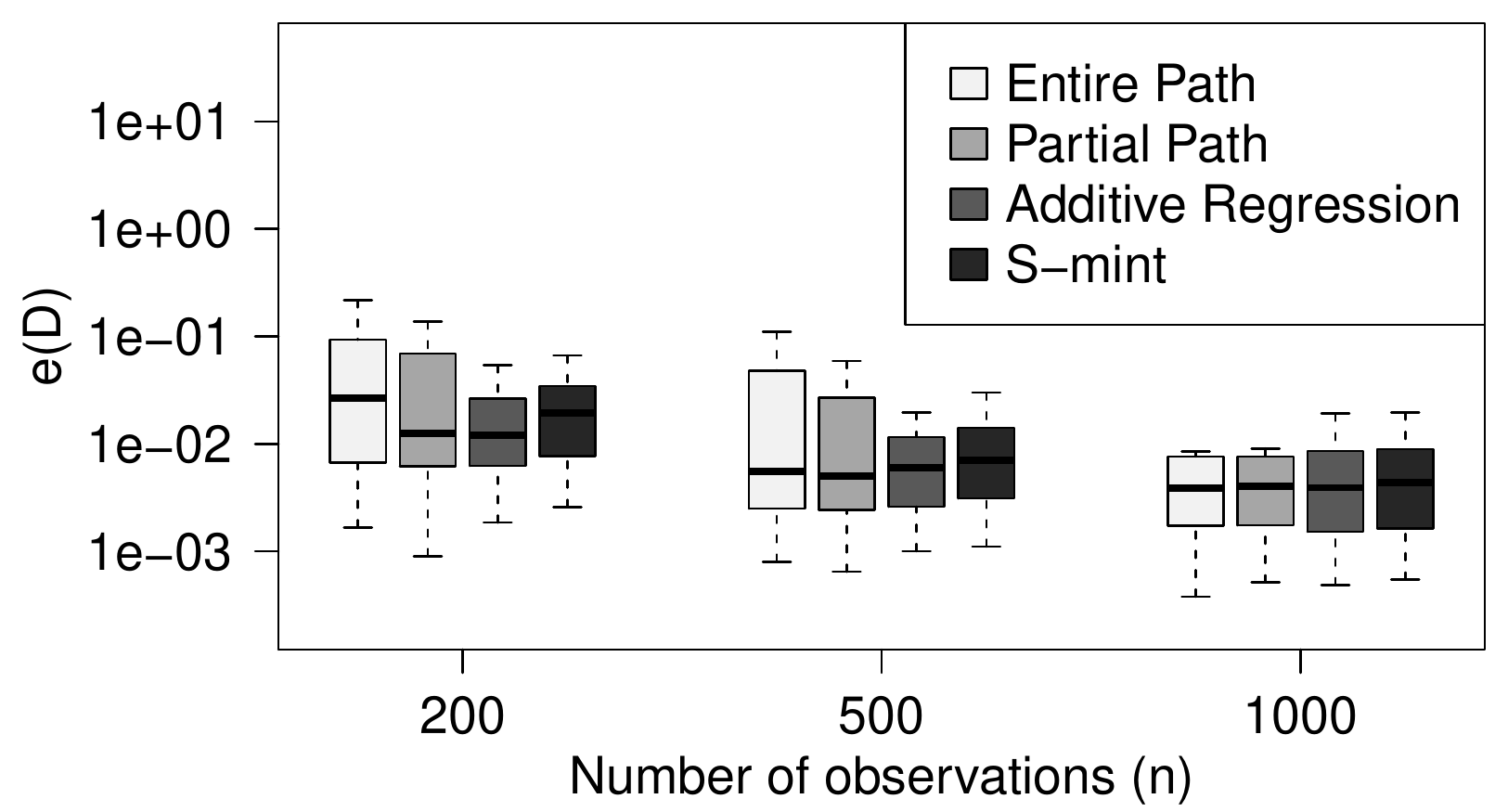}
\end{center}
\caption{Sigmoid-type additive structural equation models. Relative squared error performance 
  as in (\ref{eq:approxError}), for different numbers of 
  observations ($n$), computed on graphs that have been estimated using the
  CAM method \citep{pbjoja13}. The algorithm has been applied without
  the pruning step (left) and with the pruning step (right). We use the
  estimated parental sets as adjustment sets and the number of variables is
  $p=20$. The \emph{S-mint} regression corresponds to \emph{est S-mint} as
  described in Section \ref{subsec.twostage}.}
\label{fig:CEestimDAG}
\end{figure}

After having estimated a DAG $\hat{D}$ with the above procedure, we
randomly select $10 = |\mathcal{L}|$ index pairs $(k,j)$ such that there exists a directed
path from $X_k$ to $X_j$ in the true DAG $D^0$ and approximate the total causal
effect $\EE[X_j|\Do(X_k)]$ based on the estimated graph
$\hat{D}$. Figure~\ref{fig:CEestimDAG} displays the relative squared errors as defined in \eqref{eq:approxError}.  

All four methods show a similar performance with respect to relative
squared error on the DAGs that are obtained applying the CAM method without
feature selection. These DAGs mainly 
represent the causal order of the variables but otherwise are densely
connected. An incorrectly specified order of the variables (e.g. for small
sample sizes $n$) seems to comparably affect the \emph{S-mint} and additive regression with parental sets and the
path-based methods. If the sample size increases, the estimated graph
$\hat{D}$ is closer to the true graph $D^0$ 
which improves the estimation accuracy of causal effects for all the four methods.

\begin{figure}[h]
\begin{center}
\includegraphics[width=0.48\textwidth]{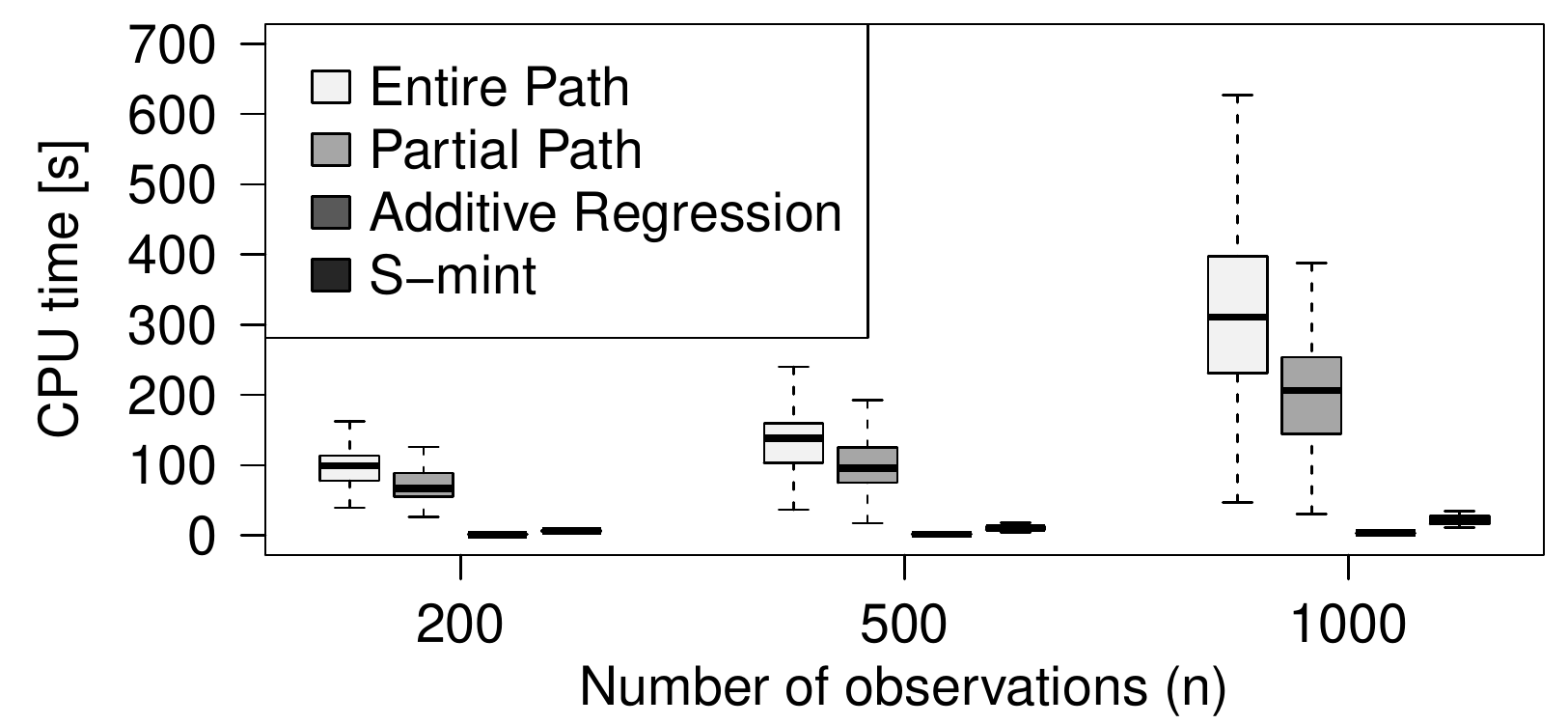} \hfill
\includegraphics[width=0.48\textwidth]{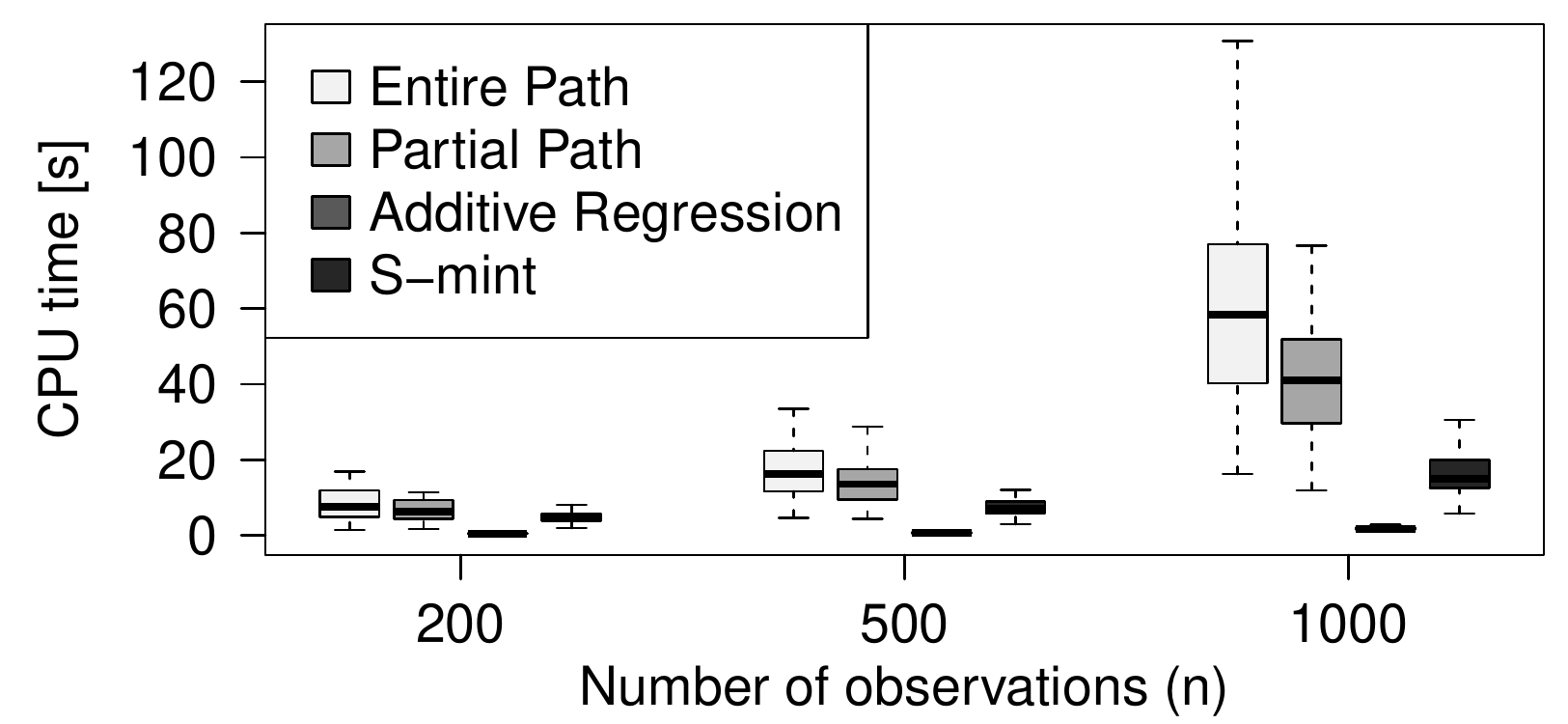} \hfill
\end{center}
\caption{Sigmoid-type additive structural equation models. CPU time performance for $n=500$ for $N=50$ graphs of $p=20$
  variables that have been estimated using the CAM method
  \citep{pbjoja13} with and without pruning step. Pruning the DAG yields
  vast computational savings for the two path-based methods. \emph{S-mint}
  and additive regression are barely affected by the pruning step and are
  considerably faster than the two path-based methods in both scenarios.}  
\label{fig:CEestimDAG.CPU}
\end{figure}

The two path-based methods approximate the causal effects more accurately
on the DAGs that are obtained without feature selection, that is, pruning
the DAG 
is not advantageous for the estimation accuracy of
causal effects, at least for a small number of observations. However, the
pruning step yields vast computational savings for the two path-based
methods as demonstrated in Figure~\ref{fig:CEestimDAG.CPU}. The
\emph{S-mint} regression is very fast in both settings and pruning the DAG
before estimating the causal effects only has a minor effect on the time
consumption and estimation accuracy. 

\subsection{Summary of the empirical results, and the advantage of the
  proposed two-stage \emph{est S-mint}
  method}\label{subsec.summaryemp}

With respect to statistical accuracy, measured with the relative squared error as in (\ref{eq:approxError}), 
we find that \emph{S-mint} and additive regression are
substantially more robust against incorrectness of the true underlying DAG
(or against a wrong order of the variables) and against model
misspecification, in comparison to the alternative path-based methods. 
The latter robustness of \emph{S-mint} is rigorously backed-up by our 
theory in Theorem~\ref{th1} and Corollary~\ref{cor1} whereas the former
seems to be due to the higher degree of localness as described in
Section~\ref{subsec.local}. As a consequence, the proposed two-stage \emph{est S-mint} (Section~\ref{subsec.twostage}) where we first 
estimate the order of the variables or the structure of the DAG (or the
Markov equivalence class of DAGs) and subsequently perform \emph{S-mint}
is expected in general to lead to reasonably accurate results (which are 
empirically quantified above for some settings). Only when the DAG is
perfectly known and the model correctly specified (here by an additive
structural equation model), which is a 
rather unrealistic assumption for practical applications, the path-based
methods were found to have a slight advantage. Thus, we recover here a
typical robustness phenomenon against model misspecification of our
nonparametric and more ``model-free'' \emph{S-mint} regression procedure. 

Our empirical findings support the use of \emph{est S-mint}, namely the combination of a structured nonparametric (or parametric) approach for estimating the DAG (or its equivalence class) in the first stage and using the robust and fully nonparametric \emph{S-mint} procedure in the second stage. The second stage leads to a  clear gain in robustness whereas the efficiency loss in case of correctly specified models is marginal or even minimal.

Regarding computational efficiency, \emph{S-mint} and in particular also
the additive regression approximation are massively faster than the
path-based procedures making them feasible 
for larger scales where the number of variables is in the thousands.

\section{Real data application}\label{sec.realdata}

In this section we want to provide two examples for the application of our
methodology to real data. We use gene expression data from the isoprenoid
biosynthesis in \textit{Arabidopsis thaliana} \citep{wille04}. The
data consists of $n=118$ gene expression measurements from $p=39$
genes. In the original work the authors try to infer connections between
the individual genes in the network using Gaussian graphical modeling. Our
goal is to find the strongest causal connections between the individual
genes. We do not standardize the original data but adjust the bandwidths in
\emph{S-mint} by scaling with the standard deviations of the corresponding
variables.

\subsection{Estimation and error control for causal connections between 
and within the pathways}

We first turn our attention to the whole isoprenoid biosynthesis dataset and want to find the causal effects within and between the
different pathways, with an error control for false positive
selections. To be able to compute the causal effects we have to
estimate a causal network. In order to do that we use the CAM method
\citep{pbjoja13}. 

We estimate a DAG using CAM with the default settings. We then
apply the \emph{S-mint}  procedure with parental sets obtained from the estimated DAG (which corresponds to the \emph{est S-mint} procedure from Section~\ref{subsec.twostage}) to rank the total causal 
effects according to their strength. 
We define the relative causal strength
$\text{CS}^{\text{rel}}_{k\rightarrow j}$ of an intervention $X_j|\Do(X_k)$
as a sum of relative distances of observational and interventional
expectation for different intervention values divided by the range of the
intervention values, i.e.  
$$
	\text{CS}^{\text{rel}}_{k\rightarrow j} = \frac{1}{R_k(d)} \sum\limits_{i=1}^{9} \frac{ |\EE[X_j] - \EE[X_j|\Do(X_k=d_i)] |}{| \EE[X_j]|},  
$$  
where we choose $d_1(X_k),...,d_9(X_k)$ to be the nine deciles of $X_k$ and we denote
their range by $R_k(d) = d_9(X_k) - d_1(X_k)$.  

\begin{figure} [h]
\begin{center}
\includegraphics[width=0.98\textwidth]{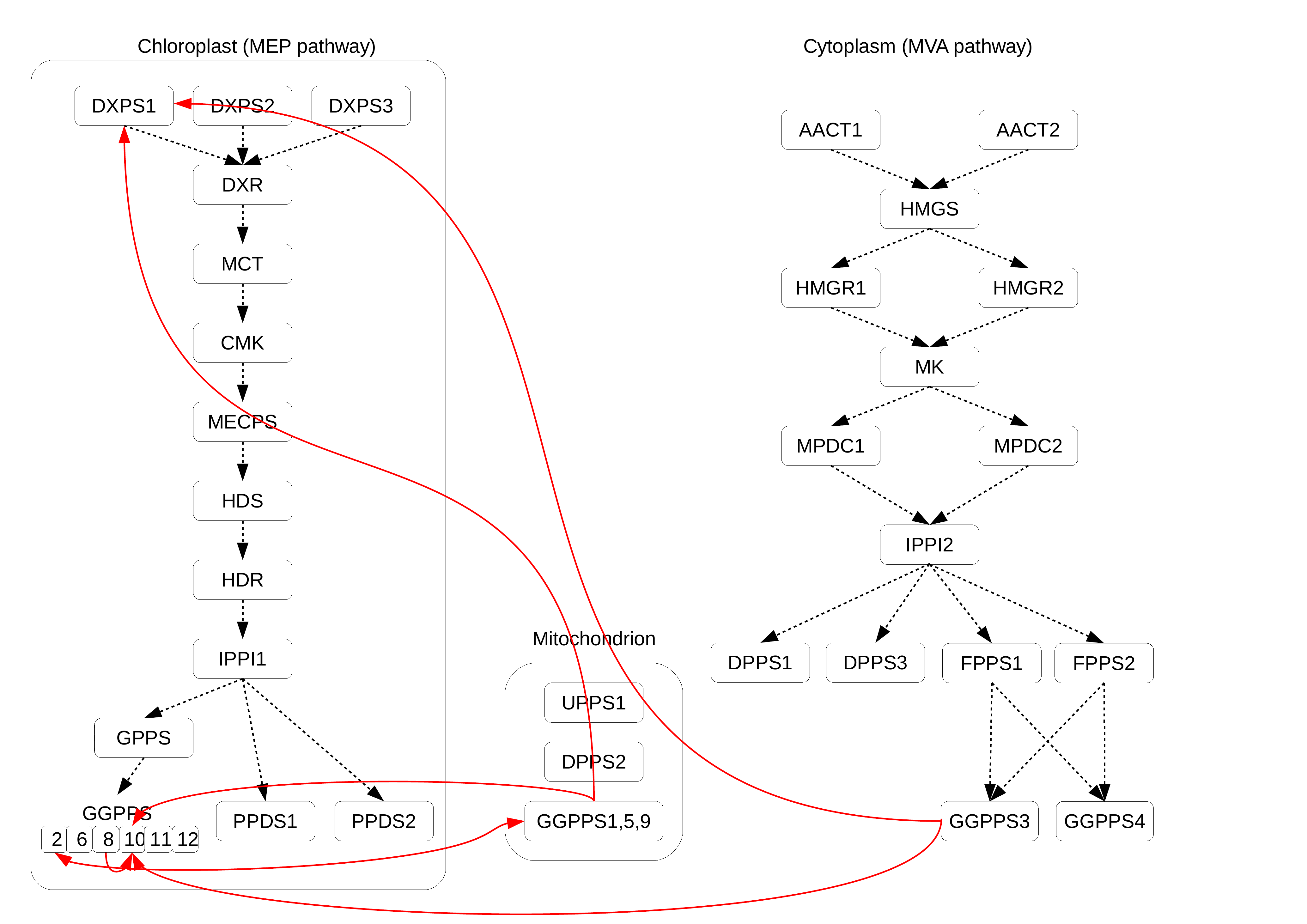}
\end{center}
\caption{Stable edges (with stability selection) for the
  \textit{Arabidopsis thaliana} dataset. The dotted arcs represent the
  metabolic network, the red solid arcs the stable total causal effects
  found by the \emph{est S-mint} method.} 
\label{fig:CErealData}
\end{figure}

To control the number of false positives (i.e. falsely
selected strong causal effects) we use stability selection
\citep{mebu10} which provides (conservative) error control under a 
  so-called (and uncheckable) exchangeability condition. We randomly
  select $100$ subsamples of size $n/2 = 59$ and repeat the procedure above
  $100$ times. For each run, we record the indices of the top $30$ ranked
  causal strengths. At the end 
we keep all index pairs that have been selected at least $66$ times in the
$100$ runs as this leads to an expected number of falsely selected edges
(false positives) which is less or equal to $2$ \citep{mebu10}. The
graphical representation of the network in Figure~\ref{fig:CErealData} is
based on \cite{wille04}.
The dotted arcs represent the underlying metabolic network (known from
biology), the six red solid arcs correspond to the 
stable index pairs found by \emph{est S-mint} with stability selection.  

None of the stable edges are opposite to the causal direction of the
metabolic network. In particular, 
we found strong total causal
effects between GGPPS variables in the MEP pathway, MVA pathway and
mitochondrion. Note that in this section we heavily rely on model 
assumption \eqref{addSEM} as the CAM method for estimating a DAG
assumes additivity of the parents. Therefore we cannot fully exploit the
advantage of the \emph{S-mint} method that it works for arbitrary
non-additive models \eqref{SEM} (but we would hope to be somewhat less
sensitive to model misspecification than with path-based methods, see for
example Figures~\ref{fig:CEmodifDAG}~and~\ref{fig:CEmodifDAG.GAM}). 

\subsection{Estimation and error control of strong causal connections within the MEP pathway} \label{ssec:realKnownOrder}

We now want to present a possible way of exploiting the very general model
assumptions of \emph{S-mint}. If the underlying order and an approximate graph structure are
known \textit{a priori}, we can use this information to proceed with
\emph{S-mint} using the order information as described in
Corollary~\ref{cor1}. 
This
relieves us from any model assumptions on the functional connections
between two variables (e.g. linearity, additivity, etc.).

\begin{figure}[!htb]
\begin{center}
\includegraphics[width=0.58\textwidth]{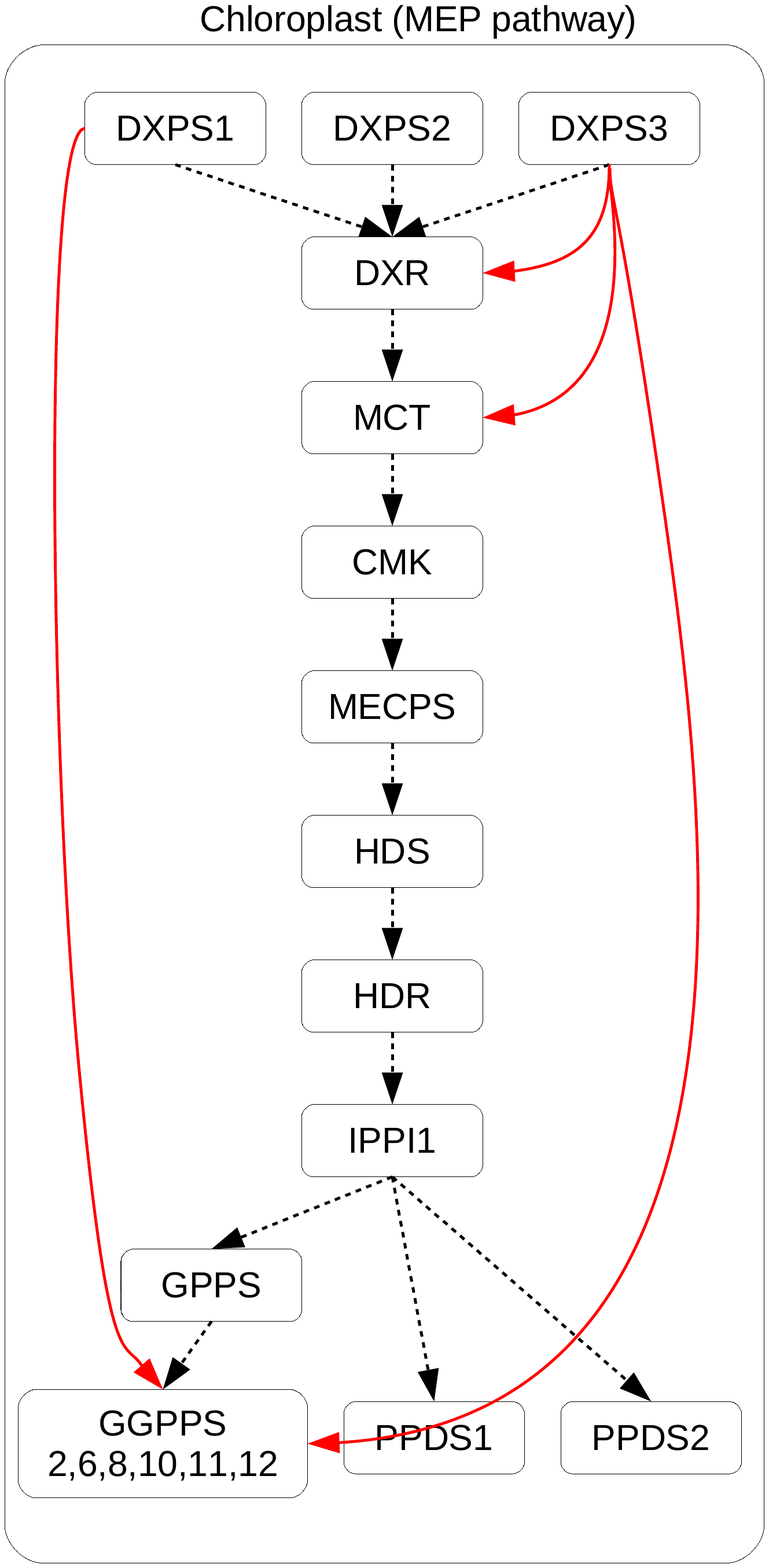}
\end{center}
\caption{Stable edges (with stability selection) for the MEP pathway in the
  \textit{Arabidopsis thaliana} dataset. The dotted arcs represent the
  metabolic network whereas the red solid arcs denote the top ranked causal
  effects found by \emph{S-mint} with adjustment sets chosen from
  the order of the metabolic network structure by considering all ancestors up to three levels back.}  
\label{fig:CErealData2}
\end{figure}

To give an example, let us focus on the genes in the MEP pathway (black box
in Figure~\ref{fig:CErealData}). The goal is to find the strongest total causal
effects within this pathway. The metabolic network (dotted arcs) is
providing us with an order of the variables which we use for \emph{S-mint}
regression as follows: 
we choose the adjustment set $S(j_X)$ in (\ref{supset.ord1}) by going three
levels back ($p_{\max} = 3$) in the causal order (to achieve a reasonably
sized set), for example, the adjustment set for CMK is {DXPS1, DXPS2,
  DXPS3, DXR, MCT}, whereas the adjustment set for GPPS is {HDS, HDR,
  IPPI1}. We cannot use the 
full set of all ancestors because there are only $n/2=59$ data points to
fit the nonparametric additive regression and marginal integration, as we
again use stability selection based on subsampling for controlling false
positive selections as described in the previous section. For each
among the $100$ subsampling runs we record the top $10$ ranked 
index pairs and keep the ones that are selected at least $65$ times out of
$100$ repetitions. This results in an expected number of false positives
being less than~$1$ \citep{mebu10}. The stable edges are shown in
Figure~\ref{fig:CErealData2}. One of the four edges corresponds to an edge
in the metabolic pathway. We find that the upper part of the pathway
contains the strongest total causal effects and therefore may be an
interesting target for intervention experiments.  

\section{Conclusions}

We considered the problem of estimating expected values of intervention
distributions, also known as total causal effects, from  
observational data. A first main result (Theorem~\ref{th1} and
Corollary~\ref{cor1}) says that if we know the local parental variables or
a superset thereof (e.g., from the order of the variables), 
there is no need to base estimation and computations on a causal graph. In fact, we can
directly infer the expected values of 
single-intervention distributions via marginal integration: we call the
procedure \emph{S-mint}. This result holds for any nonlinear
and non-additive structural equation model apart from 
mild smoothness
and regularity conditions. Hence, from another point of
view, \emph{S-mint} estimation of 
expected values of single intervention distributions is a fully
nonparametric technique and thus robust against
model misspecification of the functional form of the structural
equations. We propose an $L_2$-boosting approach for \emph{S-mint} which is
easy to use without complicated tuning of parameters and yields good
empirical results.   

We complement the robustness view-point by empirical results indicating
that \emph{S-mint} also works reasonably well when the DAG- or
order-structure is misspecified to a certain extent, as it will be the case
when we estimate these quantities from data; in fact, \emph{S-mint}
regression is substantially more robust than methods which follow all
directed paths in the DAG to infer causal effects. This suggests that the
two-stage \emph{est S-mint} procedure is most reliable for causal inference
from observational data: first estimate the DAG- or order-structure (or 
equivalence classes thereof) and second, subsequently pursue \emph{S-mint}
regression. In addition, such a procedure is computationally much faster 
than methods which exploit directed paths in (estimated) DAGs.

\paragraph{Acknowledgments.} We sincerely thank Linbo Wang and Mathias
Drton for having pointed out a major error in an earlier and very different
version of the paper. We also thank anonymous reviewers for constructive
comments. 

\bibliographystyle{apalike} 
\bibliography{reference}

\end{document}